\documentclass[12pt,draft]{article}

\usepackage{amsthm,amssymb,amsmath,amscd,amstext,mathrsfs}
\usepackage{latexsym,mathptmx,xypic}

\newtheorem{definition}{Definition}[section]
\newtheorem{theorem}{Theorem}[section]
\newtheorem{lemma}{Lemma}[section]
\newtheorem*{remark}{{\it Remark}}

\newcommand{\nc}{\newcommand}

\nc{\N}{{\mathbb N}}
\nc{\Z}{{\mathbb Z}}
\nc{\T}{{\mathbb T}}
\nc{\R}{{\mathbb R}}
\nc{\C}{{\mathbb C}}
\nc{\HH}{{\mathbb H}}

\nc{\dd}{{\rm d}}
\nc{\DD}{{\rm D}}

\nc{\ca}{{\mathscr A}}
\nc{\cb}{{\mathscr B}}
\nc{\cg}{{\mathscr G}}
\nc{\ch}{{\mathscr H}}
\nc{\cm}{{\mathscr M}}
\nc{\csu}{{\mathscr S}{\mathscr U}(2)}

\begin{document}

\title{The four dimensional Yang--Mills partition function\\ 
in the vicinity of the vacuum}

\author{G\'abor Etesi\\
\small{{\it Department of Geometry, Institute of Mathematics,}}\\
\small{{\it Budapest University of Technology and Economics,}}\\
\small{{\it M\H uegyetem rkp. 3., H-1111 Budapest, Hungary}}
\footnote{E-mail: {\tt etesi@math.bme.hu}}}

\maketitle

\pagestyle{myheadings}
\markright{G. Etesi: The Yang--Mills partition function in the vicinity of 
the vacuum}

\thispagestyle{empty}

\begin{abstract} 
The partition function of four dimensional Euclidean, 
non-supersymmetric ${\rm SU}(2)$ Yang--Mills theory is calculated in the  
perturbative and weak coupling regime i.e. in a small open ball 
about the flat connection and when the gauge coupling constant acquires a 
small but finite value. 

The computation is based on various known inequalities, valid only in four 
dimensions, providing two-sided estimates for the exponentiated 
Yang--Mills action in terms of the $L^2$-norm of the derivative of the 
gauge potential only; these estimates then give rise to Gau\ss ian-like 
infinite dimensional formal integrals involving the Laplacian hence can be 
computed via zeta-function and heat kernel techniques. It then 
turns out that these formal integrals give a sharp value for the partition 
function in the aforementioned perturbative and weak coupling regime of 
the theory. 

In the resulting expression for the partition function the original 
classical value of the coupling constant is shifted to a smaller one 
which can be interpreted as the manifestation, in this approach, of 
a non-trivial $\beta$-function and asymptotic freedom in pure non-Abelian 
gauge theories.
\end{abstract}

\centerline{AMS Classification: Primary: 81T13; Secondary: 81Q30, 
57M50, 35K08}
\centerline{Keywords: {\it Non-supersymmetric Yang--Mills partition 
function;}} 
\centerline{\it {Zeta-function regularization; Heat kernel; Asymptotic 
freedom}}


\section{Introduction and summary}
\label{one}


Computing the partition function is a central problem of Yang--Mills 
theory. For in Feynman's path integral quantization framework it is 
intrinsically equivalent with the highly non-trivial task of taking summation 
over all vacuum Feynman graphs, the computation of the partition function is 
the first and most difficult step towards the construction of the underlying 
relativistic quantum field theory. In the exposition of the problem 
mainly found in physicist's textbooks (cf. e.g. \cite{che-li,itz-zub,kak,wei}) 
the difficulties are usually attributed to the presence of a huge (namely 
gauge) symmetry of the theory alone; however the troubles have certainly much 
deeper roots related e.g. with our problematic 18-$19^{\rm th}$ century 
concept of the continuum \cite{bae,wey} and the non-existence of a good 
measure theory in infinite dimensions \cite{joh-lap}, too. Nevertheless, 
because of its central importance, permanent efforts have been made to 
calculate the partition function during the past decades. These are based 
upon taming the partition function in order to increase its computational 
accessibility by using either discretization i.e. lattice methods 
(e.g. \cite{bal1,bal2}) or yet working with the continuum but introducing 
additional structures. Very roughly speaking these latter approaches hit the 
field in three powerful waves: in the 1970-1980's various 
{\it supersymmetric} and {\it higher dimensional extensions} of pure 
Yang--Mills theory have been 
introduced making it possible to calculate their corresponding partition 
functions via Atiyah--Bott-like localization techniques, cf. 
\cite{pes-zab} (especially \cite[Chapter 10]{pes-zab}). 
Then {\it topological twisting}, an additional modification was introduced by 
Witten \cite{wit1} which together with many other ideas such as the 
Chern--Simons and conformal field theory correspondence and various 
duality conjectures, etc. led in the 1990's to revolutionary 
discoveries connecting quantum field theories and low dimensional 
differential topology \cite{wit1,wit2,sei-wit} thereby clearly 
demonstrating the indeed deep, not only physical but even mathematical, 
relevance of Yang--Mills partition functions. However, eventually together 
with Nekrasov's {\it $\Omega$-deformation} approach \cite{nek} from the early 
2000's, these supersymmetric twisting and deformation techniques, as a price 
for computability, gradually converted the Yang--Mills partition 
function, an originally certainly highly analytical object, into a rather 
purely combinatorial structure; in this way at least in part having covered or 
mixed the original physical content of Yang--Mills theory with auxiliary 
mathematical structures. 

In this paper, as a continuation of our earlier work on the Abelian 
case \cite{ete-nag}, we make an attempt to return to the original setup and 
compute the partition function of the non-supersymmetric, non-twisted, 
etc. but surely non-Abelian four dimensional Euclidean pure 
(i.e. without fermions and scalars) gauge theory. The sacrifice we make for 
not using any supersymmetric, etc. support is that unfortunately we 
shall neglect all non-perturbative (like instanton, etc.) effects which are 
however certainly key features of non-Abelian gauge thoeries; that is we shall 
consider the perturbative regime only. It is worth briefly mentioning here that 
part of our approach which in our opinion is the most interesting (and 
well-known) because works only in four dimensions. The curvature of a 
connection $\nabla =\dd +A$ looks like $F_\nabla =\dd A+A\wedge A$ i.e. 
consists of a derivative (dynamical) $\dd A$ and a quadratic (interacting) term 
$A\wedge A$ of the gauge potential. In four dimensions there is a delicate 
balance between these terms as a consequence of the Sobolev embedding 
$L^2_1\subset L^4$ which is on the borderline in four 
dimensions. Indeed, this embedding allows one to compare the $L^2$-norm of 
the $\dd A$ and $A\wedge A$ terms. Phyisically speaking this 
means that precisely in four dimensions the energy content in the Yang--Mills 
field strength is equally distributed between its dynamical and interacting 
terms.\footnote{One is tempted to say that although in dimensions different 
from four classical Yang--Mills theory can be formulated, its underlying 
quantum theory will be governed by 
$\dd A$ or $A\wedge A$ alone; hence it exhibits a different, perhaps less 
complex, behaviour.} From the mathematical aspect the existence of 
$L^2_1\subset L^4$ allows one to estimate the $L^2$-norm of the curvature of a 
connection from both below and above by various, at most quartic, 
expressions involving the $L^2$-norm of the derivative part of the 
gauge potential alone. These estimates can be re-written as Gau\ss ian-like 
expressions for the Laplacian hence can be formally Feynman integrated 
using $\zeta$-function and heat kernel techniques providing a two-sided 
estimate for the partition function. After adjusting the physical and 
technical parameters involved in this procedure, this ``scissor'' about 
the partition function closes up giving rise to an expression for it. 

For clarity we emphasize that our forthcoming calculations and assertions 
are supposed to be mathematically rigorous {\it except} precisely the 
mathematical definition of Feynman integration itself (which of course 
is a crucial point); this latter thing will be rather treated only formally 
throughout the text but in the standard way by using $\zeta$-function 
regularization. We also emphasize that what we are going to write throughout 
the text as 
\[Z_\varepsilon(\R^4,\tau)\] 
and want to calculate is {\it not} an approximation 
of the full partition function $Z(\R^4,\tau)$ of four dimensional 
non-supersymmetric Yang--Mills theory 
(containing all instanton and other non-perturbative contributions) but a 
contribution of the {\it vicinity of the vacuum} i.e. the 
complete perturbative regime in the weak coupling limit 
to the full partition function. Of course an important 
question is whether or not $Z_\varepsilon(\R^4,\tau)$ already gives rise to 
the leading contribution to $Z(\R^4,\tau)$ i.e. whether or not by some (hidden) 
localization mechanism already $Z(\R^4,\tau)\approx Z_\varepsilon(\R^4,\tau)$. 
The answer for this question is certainly negative because on the one hand 
localization phenomena are expected to occur only in supersymmetrized 
Yang--Mills theories \cite{pes-zab} (and we are not dealing with them here) and 
on the other hand instantons with non-zero topological numbers surely give 
further relevant contributions to the full partition function $Z(\R^4,\tau)$ 
hopefully rendering it a nice modular form in its (probably quantum 
corrected) $\tau\in\C^+$ variable as indicated by various $S$-duality 
conjectures (far from being complete cf. e.g. \cite{oli-wit,sen,vaf-wit,wit3}). 
Nevertheless $Z_\varepsilon(\R^4,\tau)$ already alone is expected to 
reveal something from the quantum behaviour of gauge theory. 

After these careful circumscriptions, limitations and clarifications 
our main {\it formal} result can be summarized as follows. For the very 
technical details we refer to Sections \ref{three} and \ref{four} below.  
\begin{theorem}
Consider a non-supersymmetric pure ${\rm SU}(2)$ gauge theory with 
complex coupling constant $\tau\in\C^+$ over the Euclidean $4$-space 
$(\R^4,\eta)$. Take a constant $0<\varepsilon<\sqrt{8}\pi$ and 
consider those ${\rm SU}(2)$ connections $\nabla$ which are close to 
the flat connection $\nabla^0$ in the sense that 
$\Vert F_\nabla\Vert_{L^2(\R^4)}<\varepsilon$. Let 
$Z_{\varepsilon}(\R^4,\tau)$ denote the corresponding {\em 
truncated partition function} of the theory obtained by formally Feynman 
integrating the exponentiated Yang--Mills action over gauge 
equivalence classes of ${\rm SU}(2)$ connections close to the flat connection 
against a formal measure provided by the round sphere 
$(S^4,g_R)$ of radius $R$ which is a one-point conformal compactification 
of $(\R^4,\eta)$ (hence this formal measure and thus 
$Z_\varepsilon (\R^4,\tau)$ itself may in principle depend on $R$). 

Provided the complex coupling constant $\tau\in\C^+$ has large enough imaginary
part (the weak coupling regime) and accordingly both the vicinity 
parameter $\varepsilon$ is small enough (the perturbative regime) and 
the compactification radius $R$ is small enough (a technical condition on the 
formal measure) then, using $\zeta$-function regularization and heat kernel 
techniques, the truncated partition function can be computed and 
\[Z_{\varepsilon}(\R^4,\tau)=\left(\frac{{\rm Im}\tau}{2\pi^2N^2}
\right)^{-\frac{11}{20}}
\frac{2^{-\frac{11}{20}}}{\sqrt{\pi}}
\cos\begin{smallmatrix}\left(\frac{11\pi}{40}\right)\end{smallmatrix}
\Gamma\begin{smallmatrix}\left(\frac{9}{40}\right)\end{smallmatrix}
{\rm e}^{\frac{3}{2}\zeta'_{\Delta_1}(0)-3\zeta'_{\Delta_0}(0)}\]
where $N$ is a constant satisfying $1\leqq N<\sqrt{2}$ 
and $\Gamma$ is Euler's Gamma function; morover 
$\zeta_{\Delta_k}$ are the $\zeta$-functions 
of Laplacians acting on $k$-forms over $(S^4,g_R)$.  

The truncated partition function $Z_\varepsilon 
(\R^4,\tau)$ depends on $R$ only through the formal 
determinant term 
${\rm e}^{\frac{3}{2}\zeta'_{\Delta_1}(0)-3\zeta'_{\Delta_0}(0)}$.
More precisely provided the radii $0<R_1<R_2$ are both small enough hence 
the corresponding $(S^4,g_{R_i})$ are two allowed conformal one-point 
compactifications of $(\R^4,\eta)$ then 
\[Z^1_\varepsilon(\R^4,\tau)=
\left(\frac{R_1}{R_2}\right)^{\frac{11}{10}}Z^2_\varepsilon(\R^4,\tau)\]
demonstrating that the conformal invariance of classical gauge theory 
breaks down.
\label{fotetel}
\end{theorem}  

\begin{remark}\rm 1. 
${\rm e}^{\frac{3}{2}\zeta'_{\Delta_1}(0)-3\zeta'_{\Delta_0}(0)}=10.710...$ 
over the unit sphere and this expression of the formal determinant 
can be further expanded in terms of the derivatives of the 
standard Riemann and Hurwitz $\zeta$-functions (cf. e.g. 
\cite{eli-lyg-vas,kum,qui-cho}); however the result is not 
promising hence omitted. One might hope to obtain nicer determinant 
expressions by introducing Dirac fermions into the theory, too. 
Also cf. \cite{cal-tau}.  

2. The particular numerical values of the determinant above, the exponent 
$-\frac{11}{20}$ or the coefficient $N$ in $Z_\varepsilon(\R^4,\tau)$ 
bear no direct physical meaning for they depend on the particular 
regularization scheme used to make sense of infinite dimensional 
integrals here. Concerning $N$ it is essentially nothing else than a 
good choice for a constant in Uhlenbeck's gauge fixing theorem \cite{uhl2} 
(see Lemma \ref{kettosegyenlotlenseglemma} below) and the only relevant point 
is that $N<\sqrt{2}$ must hold in order our method to work (see Lemma 
\ref{eredmeny}). This is provided by the at least one universal property of 
$N$ namely that whatever its value is, it is conformally invariant and surely 
$1\leqq N$ such that $N\rightarrow 1$ as $\varepsilon\rightarrow 0$ 
(see Lemma \ref{kettosegyenlotlenseglemma}).
 
3. Nevertheless Theorem \ref{fotetel}, when compared with the analogous 
Abelian result, admits an interesting physical interpretation in the context of 
{\it asymptotic freedom} which is a key property 
of non-Abelian gauge theories. The complex coupling constant is defined as 
$\tau :=\frac{\theta}{2\pi}+\frac{4\pi}{e^2}\sqrt{-1}$ 
where $\theta$ is the so-called $\theta$-parameter and $e$ is the coupling 
constant of the gauge theory. It enters the theory 
at its classical level i.e. $\tau$ appears already in its defining 
action. However it is well-known that in a non-supersymmetric four dimensional 
gauge theory, meanwhile $\theta$ is unaffected hence is a true quantum 
parameter, $e$ is subject to quantum corrections i.e. the theory has a 
{\it non-trivial $\beta$-function}. Therefore in our case it is intriguing 
to physically interpret the appearance of the purely technical-mathematical 
constant $N$ in Theorem \ref{fotetel} as a {\it quantum correction} of the 
classical gauge coupling. That is, by recalling from \cite{ete-nag} the 
full partition function over $(S^4,g_R)$ in the ${\rm U}(1)$ case: 
\[Z(\R^4,\tau)=
\left(\frac{{\rm Im}\tau}{8\pi^2}\right)^{-\frac{11}{60}}
{\rm e}^{\frac{1}{2}\zeta'_{\Delta_1}(0)-\zeta'_{\Delta_0}(0)}\] 
we cannot resist the temptation to re-write the truncated ${\rm SU}(2)$ 
partition function computed here as 
\[Z_{\varepsilon}(\R^4,\tau)=\left(\frac{{\rm Im}\tau_{\rm eff}}{8\pi^2}
\right)^{-\frac{11}{20}}
\frac{2^{-\frac{11}{20}}}{\sqrt{\pi}}
\cos\begin{smallmatrix}\left(\frac{11\pi}{40}\right)\end{smallmatrix}
\Gamma\begin{smallmatrix}\left(\frac{9}{40}\right)\end{smallmatrix}
{\rm e}^{\frac{3}{2}\zeta'_{\Delta_1}(0)-3\zeta'_{\Delta_0}(0)}\]
i.e. absorb $N$ into the classical $\tau$ in this way shifting it to 
$\tau_{\rm eff}=\frac{\theta}{2\pi}+
\frac{4\pi}{e_{\rm eff}^2}\sqrt{-1}$ where 
$e_{\rm eff}:=\frac{N}{2}e$ is 
considered as an effective i.e., perturbatively quantum corrected coupling 
constant (the inessential numerical term $\frac{2^{-\frac{11}{20}}}{\sqrt{\pi}}
\cos\begin{smallmatrix}\left(\frac{11\pi}{40}\right)\end{smallmatrix}
\Gamma\begin{smallmatrix}\left(\frac{9}{40}\right)\end{smallmatrix}=1.013...$ 
rather looks like a non-Abelian correction to the formal determiant). 
However the key property of $N$ i.e. that $1\leqq N<\sqrt{2}$ makes 
sure that $e_{\rm eff}<e$ 
rendering the effective gauge coupling constant {\it smaller} than its 
classical value. This is qualitatively consistent with our picture on 
asymptotic freedom in pure non-Abelian gauge theories, the net effect of a 
highly counter-intuitive Yang--Mills-charge-anti-screening-mechanism 
generated by virtual charged gauge bosons floating around the real ones. 
In addition it is well-known (cf. e.g. \cite{itz-zub}) that the presence of a 
non-trivial $\beta$-function in Yang--Mills theory is in conjunction with the 
breakdown of its classical conformal symmetry at the quantum level introduced 
by the formal integration measure lacking conformal invariance; 
hence our physical interpretation of Theorem \ref{fotetel} is consistent from 
this angle as well. 

4. We can also make a comment regarding $S$-duality \cite{vaf-wit}. In 
Theorem \ref{fotetel} it is assumed that $\tau$ has large (but finite!) 
imaginary part that is, the gauge coupling $e$ is small. This assumption is 
physically clear because in this weak coupling regime the existence of 
convergent perturbation series is reasonable. The weak and the strong 
coupling regimes of a gauge theory 
are related by $S$-duality transformations. Supposing that $\tau_{\rm eff}$ is 
already meaningful at the quantum level, more precisely after taking into 
account at least small perturbative quantum corrections as in 
Theorem \ref{fotetel} and recalling the identity 
${\rm Im}\left(-\frac{1}{\tau_{\rm eff}}\right)=
\frac{1}{\tau_{\rm eff}\:\overline{\tau}_{\rm eff}}\:{\rm Im}\tau_{\rm eff}$ 
we recognize that the truncated partition function is a modular form 
with (holomorphic and anti-holomorphic) weight $(\frac{11}{20},\frac{11}{20})$ 
hence $Z_{\varepsilon}(\R^4,\tau_{\rm eff})$ has a promising behaviour under 
$S$-duality transformations \cite{vaf-wit}. Of course to say something more 
definitive on this topic (for instance what about the modular properties of 
the full partition function with some meaningful $\tau_{\rm eff}$ and how 
${\rm SU}(2)$ is replaced with its Langlands dual group ${\rm SO}(3)$, etc.) 
one would need to calculate the complete partitition function 
$Z(\R^4,\tau_{\rm eff})$ consisting of all instanton, etc. corrections; 
this is however far beyond our technical skills at this stage of the art. 

5. Finally for future work we record here without proof that essentially by 
verbatim repeating the calculations below the partition function can also be 
computed in the {\it vicinity of an (anti-)instanton} $\nabla^k$ with instanton 
number $k\in\Z$ as well. It takes the shape 
${\rm e}^{-\sqrt{-1}\pi k\tau}Z_{\varepsilon,k}(\R^4,\tau)$ if $k\geqq 0$ or 
similarly ${\rm e}^{\sqrt{-1}\pi k\overline{\tau}}Z_{\varepsilon, 
k}(\R^4,\tau)$ if $k\leqq 0$ where $Z_{\varepsilon, k}(\R^4,\tau)$ is 
an expression analogous to $Z_{\varepsilon, 0}(\R^4,\tau):=Z_{\varepsilon} 
(\R^4,\tau )$ in Theorem \ref{fotetel} such that the various ordinary 
Laplacians $\Delta_i=\dd\dd^*+\dd^*\dd$ and their corresponding functions 
$\zeta_{\Delta_i}$ are to be replaced with the twisted ones $\Delta^k_i:=
\dd_{\nabla^k}\dd^*_{\nabla^k}+\dd^*_{\nabla^k}\dd_{\nabla^k}$ and 
$\zeta_{\Delta^k_i}$ respectively. However even knowing these further 
contributions from instanton vicinities we still cannot {\it a priori} 
conclude that the full partition function would be a sum of these terms only. 
\end{remark}

\noindent The paper is organized as follows. In Section \ref{two} we 
recall the calculation of the quadratic Gau\ss ian and certain quartic 
Gau\ss ian integrals in finite dimensions. The computation of these latter 
integrals is due to Svensson \cite{sve}. The resulting formulata allow formal 
generalizations to infinite dimensions. Then in Section \ref{three} classical 
pure gauge theory with $\theta$-term is introduced in the standard way and its 
truncated partition function is computed by evaluating these infinite 
dimensional formal integrals using $\zeta$-function and heat kernel techniques. 
Finally, Section \ref{four} is an Appendix and consists a well-known no-go 
result from infinite dimensional measure theory \cite{gel-vil,joh-lap}. 
This has been added to gain a more comprehensive picture.


\section{Some quadratic and quartic Gau\ss ian integrals}
\label{two}


In this preliminary section we recall the computation of the well-known 
quadratic Gau\ss ian and a less-known quartic Gau\ss ian integral in finite 
dimensions; these considerations then allow us to formally generalize these 
integrals to infinite dimensions which is the relevant case 
for quantum field theory.

{\it The Gau\ss ian integral.} Let $(\R^m,\eta)$ 
be the $m$ dimensional Euclidean space and $S:\R^m\times\R^m\rightarrow\R$ a 
positive definite symmetric bilinear form on it given by $S(x,x):=
\eta(x,Mx)$ where $M:\R^m\rightarrow\R^m$ is a positive symmetric 
matrix whose eigenvalues therefore are real and satisfy 
$0<\lambda_i<+\infty$ for all $i=1,\dots,m$. Using a linear change of 
variables one can pass to a principal axis basis of $S$ i.e. in which it 
looks like $S(y,y)=\lambda_1y_1^2+\dots+\lambda_my_m^2$ and then performing a 
further change of variables $u_i:=\sqrt{\lambda_i}y_i$ we find that 
\[\lim\limits_{a_i\rightarrow+\infty}
\int\limits_{-a_i}^{+a_i}{\rm e}^{-\lambda_iy_i^2}\dd y_i=
\lim\limits_{a_i\rightarrow+\infty}\frac{1}{\sqrt{\lambda_i}}
\int\limits_{-\sqrt{\lambda_i}a_i}^{+\sqrt{\lambda_i}a_i}
{\rm e}^{-u_i^2}\dd u_i=\frac{\sqrt{\pi}}{\sqrt{\lambda_i}}\]
hence taking their product we come up with
\[\int\limits_{\R^m}{\rm e}^{-S(x,x)}\dd 
x=\prod\limits_{i=1}^m\frac{\sqrt{\pi}}{\sqrt{\lambda_i}}=
\frac{\pi^{\frac{m}{2}}}{\sqrt{\det M}}\]
giving rise to the well-known result. This integral has a truncated version, 
too. Let $0<\delta<+\infty$ be a fixed number and using an 
orthonormal frame $\{{\bf e}_1,\dots,{\bf e}_m\}$ adapted to $S$ let 
\begin{equation}
C^m_\delta:=\left\{y\in\R^m\:\left\vert\:y=\sum\limits_{i=1}^my_i{\bf e}_i\:,\:
M{\bf e}_i=\lambda_i{\bf e}_i\:,\:-\frac{\delta}{\sqrt{\lambda_i}}<y_i<
+\frac{\delta}{\sqrt{\lambda_i}}\right.\right\}
\label{hiperkocka}
\end{equation}
denote the ``principal axis hypercube'' of $S$ more precisely an open 
rectangular parallelepiped whose edges are parallel with the principal axes 
labeled by the eigenvalues $\lambda_i$ of $S$ and having sizes 
$\frac{2\delta}{\sqrt{\lambda_i}}$ respectively. Then introducing 
$a_i:=\frac{\delta}{\sqrt{\lambda_i}}$ we can repeat the previous calculation 
as follows:
\[\int\limits_{-a_i}^{+a_i}{\rm e}^{-\lambda_iy_i^2}\dd y_i=
\frac{1}{\sqrt{\lambda_i}}\int\limits_{-\delta}^
{+\delta}{\rm e}^{-u_i^2}\dd u_i=\frac{\sqrt{K(\delta)}}{\sqrt{\lambda_i}}\]
where $K(\delta)$, the square of the classical error function is defined as 
\[\sqrt{K(\delta)}:=\int\limits_{-\delta}^{+\delta}{\rm 
e}^{-u_i^2}\dd u_i =2\sum\limits_{j=0}^{+\infty}(-1)^j
\frac{\delta^{2j+1}}{j!(2j+1)}=2\left(\delta-\frac{\delta^3}{3}+
\frac{\delta^5}{10}-\frac{\delta^7}{42}+\dots\right)\:\:\:.\]
It is independent of $S$ and is monotonly increasing in 
$0\leqq\delta\leqq+\infty$ such that $0\leqq K(\delta)\leqq\pi$. Taking 
product again we obtain an expression 
\[\int\limits_{C^m_\delta}{\rm e}^{-S(x,x)}\dd 
x=\prod_{i=1}^m\frac{\sqrt{K(\delta)}}{\sqrt{\lambda_i}}=
\frac{K(\delta)^{\frac{m}{2}}}{\sqrt{\det M}}\leqq
\frac{\pi^{\frac{m}{2}}}{\sqrt{\det M}}\]
for the integral over the principal axis hypercube, similar for the entire 
integral above.

{\it A Gau\ss ian-like integral}. Now let us compute a more general 
integral following Svensson \cite{sve}. Namely, picking two positive definite 
bilinear forms $S_1,S_2$, we are interested in the quartic integral 
\[\int\limits_{\R^m}{\rm e}^{-S_1(x,x)^2-S_2(x,x)}\dd x\:\:.\]
Consider $\gamma_s:=\{u+\sqrt{-1}\:s\:\vert\: u\in\R\}\subset\C$ i.e. a 
straight line in the complex plane running parallel with the real axis 
$\R\subset\C$. Introducing $t:=\gamma_s(u)$ it is easy to see that 
$\int_{\gamma_s}{\rm e}^{-\left(\frac{t}{2}-\sqrt{-1}\:S_1(x,x)\right)^2}
\dd t$ exists such that its value is equal to 
$2\sqrt{\pi}$ hence independent of $s\in\R$. Referring to \cite{sve} we 
adjust our integral by carefully inserting the Gau\ss ian integral 
$1=\frac{1}{2\sqrt{\pi}}\int_{\gamma_s}{\rm e}^{-\left(\frac{t}{2}-\sqrt{-1}
\:S_1(x,x)\right)^2}\dd t$ as follows:
\begin{eqnarray}
\int\limits_{\R^m}{\rm e}^{-S_1(x,x)^2-S_2(x,x)}\dd x&=&
\int\limits_{\R^m}{\rm e}^{-S_1(x,x)^2-S_2(x,x)}
\left(\frac{1}{2\sqrt{\pi}}\int\limits_{\gamma_s}
{\rm e}^{-\left(\frac{t}{2}-\sqrt{-1}\:S_1(x,x)\right)^2}\dd t\right)
\dd x\nonumber\\
&=&\frac{1}{2\sqrt{\pi}}\int\limits_{\R^m}\left(\:\:
\int\limits_{\gamma_s}{\rm e}^{-S_2(x,x)+\sqrt{-1}
\:tS_1(x,x)-\frac{t^2}{4}}\dd t\right)\dd x\:\:.\nonumber
\end{eqnarray}
If $s\geqq 0$ then $\left\vert\int_{\R^m}{\rm e}^{-S_2(x,x)+\sqrt{-1}\:
tS_1(x,x)}\dd x\right\vert\leqq\int_{\R^m}{\rm e}^{-S_2(x,x)-sS_1(x,x)}
\dd x<+\infty$ for every fixed $t$ hence this integral exists. 
Moreover since the corresponding matrix $M_2-\sqrt{-1}\:tM_1$ is symmetric 
therefore diagonalizable, we can proceed in the standard way as above to get 
\begin{eqnarray}
\int\limits_{\R^m}{\rm e}^{-S_2(x,x)+\sqrt{-1}
\:tS_1(x,x)}\dd x&=&
\pi^\frac{m}{2}\big(\det(M_2-\sqrt{-1}\:tM_1)\big)^{-\frac{1}{2}}\nonumber\\
&=&(\sqrt{-1}\pi)^\frac{m}{2}\big(\det M_1\big)^{-\frac{1}{2}}
\big(\det(t{\bf 1}+\sqrt{-1}M_2M_1^{-1})\big)^{-\frac{1}{2}}\nonumber\\
&=&(\sqrt{-1}\pi)^\frac{m}{2}
\big(\det M_1\big)^{-\frac{1}{2}}
\big((t-z_1)\dots(t-z_m)\big)^{-\frac{1}{2}}\nonumber
\end{eqnarray}
where $z_1,\dots,z_m\in\C$ are the (not necessarily different) eigenvalues 
of the matrix $-\sqrt{-1}M_2M_1^{-1}$. Consequently if 
$\int_{\gamma_s}{\rm e}^{-\frac{t^2}{4}}
\big((t-z_1)\dots(t-z_m)\big)^{-\frac{1}{2}}\dd t$ also exists and is single 
valued the two integrations are interchangable via Fubini's theorem 
and we end up with
\[\int\limits_{\R^m}{\rm e}^{-S_1(x,x)^2-S_2(x,x)}\dd x=
\frac{(\sqrt{-1}\pi)^\frac{m}{2}}{2\sqrt{\pi}\sqrt{\det M_1}}
\int\limits_{\gamma_s}
\frac{{\rm e}^{-\frac{t^2}{4}}}{\sqrt{(t-z_1)\dots (t-z_m)}}\:\dd t\:\:.\]
Therefore our task is to arrange $\gamma_s$ with $s\geqq0$ so that the 
corresponding complex integral exists and is single valued. Certainly 
existence is achieved if $\gamma_s\subset\C$ does not hit 
$z_1,\dots,z_m$ (beacuse any of them might be a multiple eigenvalue hence 
might give a pole in the integrand). In order to make the integral single 
valued we perform usual branch cutting. Firstly, 
$z_1,\dots,z_m$ are clearly branching points of the
integral and if $m$ is even then these are the only branching points;
if $m$ is odd then beyond them the infinitely remote point is also a 
branching point. Secondly, $M_1^{-\frac{1}{2}}$ exists and is positive 
symmetric; since the eigenvalues of 
$M_2M_1^{-1}=(M_2M_1^{-\frac{1}{2}})M_1^{-\frac{1}{2}}$ and 
$M_1^{-\frac{1}{2}}(M_2M_1^{-\frac{1}{2}})$ coincide and the 
latter operator is positive symmetric, the eigenvalues of $M_2M_1^{-1}$ 
continue to be positive real numbers. Thus all the eigenvalues 
of $-\sqrt{-1}\:M_2M_1^{-1}$ are in fact aligned along the 
negative imaginary axis according to their magnitude i.e. we can suppose 
$0>z_1\geqq z_2\geqq\dots\geqq z_m>-\sqrt{-1}\:\infty$. Let us therefore 
do branch cutting in the standard way: cut up $\C$ along 
the at most $\big[\frac{m+1}{2}\big]$ segments of the negative imaginary axis 
connecting $z_1$ with $z_2$ (if $z_1\not=z_2$), $z_3$ with $z_4$ 
(if $z_3\not=z_4$) and finally $z_{m-1}$ with $z_m$ (if 
$z_{m-1}\not=z_m$) whenever $m$ is even; or $z_1$ with $z_2$ 
(if $z_1\not=z_2$), $z_3$ with $z_4$ (if $z_3\not=z_4$) and finally $z_m$ with 
$-\sqrt{-1}\:\infty$ whenever $m$ is odd. Thus the complex integral will 
be single valued if $\gamma_s\subset\C$ avoids these cutting 
segments as well.\footnote{Or equivalently we can lift any $\gamma_s$ not 
hitting the eigenvalues over the corresponding at most 
$\big[\frac{m+1}{2}\big]$-genus Riemann surface regarded as 
a branching cover of the Riemann sphere and then define the already 
single-valued integral there.} 
Thus to summarize, $\int_{\gamma_s}{\rm e}^{-\frac{t^2}{4}}
\big((t-z_1)\dots(t-z_m)\big)^{-\frac{1}{2}}\dd t$ 
both exists and is single valued if we take any $\gamma_s(u)=u+\sqrt{-1}\:s$ 
with $s\geqq 0$. 

Let us specialize from now on to the case $S_1:=c_1S$ and $S_2:=c_2S$ with 
$c_1,c_2>0$ real constants; this yields $-\sqrt{-1}(c_2M)(c_1M)^{-1}=
-\sqrt{-1}\frac{c_2}{c_1}{\bf 1}$ hence $z_1=\dots=z_m=
-\sqrt{-1}\frac{c_2}{c_1}\notin\R$. Therefore either there is no branch cutting 
if $m$ is even or there is a single branch cutting running from 
$-\sqrt{-1}\frac{c_2}{c_1}$ to $-\sqrt{-1}\:\infty$ 
if $m$ is odd. We eventually come up with  
\[\int\limits_{\R^m}{\rm e}^{-c^2_1S(x,x)^2-c_2S(x,x)}\dd x=
\frac{(\sqrt{-1}\pi)^{\frac{m}{2}}}{2\sqrt{\pi}\sqrt{\det (c_1M)}}
\int\limits_{\gamma_s}\left(t+\sqrt{-1}\:\frac{c_2}{c_1}\right)^{-\frac{m}{2}}
{\rm e}^{-\frac{t^2}{4}}\:\dd t\]
together with the truncated integral
\begin{equation}
\int\limits_{C^m_\delta}{\rm e}^{-c_1^2S(x,x)^2-c_2S(x,x)}\dd x=
\frac{(\sqrt{-1}\:K(\delta))^{\frac{m}{2}}}{2\sqrt{\pi}\sqrt{\det (c_1M)}}  
\int\limits_{\gamma_s}\left(t+\sqrt{-1}\:\frac{c_2}{c_1}\right)^{-\frac{m}{2}}
{\rm e}^{-\frac{t^2}{4}}\:\dd t
\label{reszlegesaltalanositottgauss}
\end{equation}
where $\gamma_s(u)=u+\sqrt{-1}\:s$ with any $s\geqq 0$ is the contour as 
before. It is easy to see that taking the 
limit $c_1\rightarrow 0$ these integrals reduce to the corresponding (i.e. 
the full or the truncated, respectively) Gau\ss ian ones. However we shall be 
more interested in the limit $c_2\rightarrow 0$ of the full 
(i.e. not-truncated) integral which readily looks like
\begin{equation}
\int\limits_{\R^m}{\rm e}^{-c^2_1S(x,x)^2}\dd x=
\frac{(\sqrt{-1}\pi)^{\frac{m}{2}}}{2\sqrt{\pi}\sqrt{\det (c_1M)}}
\int\limits_{\gamma_s}t^{-\frac{m}{2}}
{\rm e}^{-\frac{t^2}{4}}\:\dd t
\label{gauss}
\end{equation}
where now we allow $\gamma_s(u)=u+\sqrt{-1}\:s$ with $s>0$ only to avoid the 
pole at the origin (if $m>1$) as well as the single branch cutting along the 
whole non-positive imaginary axis (if $m$ is odd). 
 
Having warmed up with these rigorous but only finite dimensional results, 
let us generalize them to infinite dimensions at least formally. Let 
$(M,g)$ be a connected, compact, oriented Riemannian $4$-manifold without 
boundary and consider the Laplacian $\Delta_k:C^\infty 
(M;\wedge^kM)\rightarrow C^\infty (M;\wedge^kM)$ i.e. the second 
order linear, symmetric, elliptic partial differential operator 
$\Delta_k=\dd\dd^*+\dd^*\dd$ naturally acting on the space of smooth 
$k$-forms. This space admits Hilbert space completions like 
$L^2_s(M;\wedge^kM)$ for any $s\in\R$ and one can demonstrate via elliptic 
regularity that $\Delta_k$ extends to a densly defined, 
self-adjoint, unbounded linear operator $\Delta_k:L^2(M;\wedge^kM)
\rightarrow L^2(M;\wedge^kM)$. By elliptic regularity the kernel of this map 
contains precisely the space $\ch^k(M)\subset C^\infty (M;\wedge^kM)\subset 
L^2(M;\wedge^kM)$ of smooth harmonic $k$-forms; by the Hodge decomposition 
theorem this kernel is isomorphic to the de Rham cohomology group $H^k(M)$ 
hence is finite dimensional i.e. a closed subspace. Therefore 
$c\Delta_k$ with $c>0$ a real constant gives rise to a positive 
self-adjoint operator on the orthogonal complement Hilbert space 
\[\ch^k(M)^\perp\subset L^2(M;\wedge ^kM)\:\:.\] 
By the finite dimensional analogue (\ref{gauss}) it is therefore convenient 
to define a non-truncated quartic integral involving the Laplacian as 
\begin{equation}
\int\limits_{\ch^k(M)^\perp}{\rm e}^{-\left(a\:,\:
c\Delta_ka\right)^2_{L^2(M)}}\:\DD a
:=\frac{(\sqrt{-1}\pi)^{\frac{1}{2}{\rm rk'}(c\Delta_k)}}
{2\sqrt{\pi}\sqrt{{\rm det'}\left(c\Delta_k\right)}}\:
\int\limits_{\gamma_s}t^{-\frac{1}{2}{\rm rk'}(c\Delta_k)}
{\rm e}^{-\frac{t^2}{4}}\:\dd t
\label{nemcsavartgauss}
\end{equation}
where the regularized rank ${\rm rk'}$ and determinant ${\rm det'}$ is 
yet to be defined somehow. 

Likewise, let $C_\delta\subset L^2(M;\wedge^kM)$ be the ``principal axis 
hypercube'' for $\Delta_k$ defined as in the finite dimensional case 
(\ref{hiperkocka}) more precisely as the corresponding 
{\it finite} linear combinations of the eigen-forms of $\Delta_k$. 
Note that by elliptic regularity these eigen-forms belong to 
$C^\infty(M;\wedge^kM)\subset L^2(M;\wedge^kM)$ but in spite of the fact that 
they span a dense subspace of $L^2(M;\wedge^kM)$ the subset $C_\delta$ 
is {\it not} open (unlike in finite dimensions). This is because the 
eigenvalues of the Laplacian form an unbounded 
sequence i.e. $\lambda_i\rightarrow+\infty$ hence the size of the edges of 
$C_\delta$ satisfy $2a_i\rightarrow 0$ as $i\rightarrow+\infty$. Keeping in 
mind this subtlety and taking into account (\ref{reszlegesaltalanositottgauss}) 
nevertheless we put 
\begin{equation}
\int\limits_{C_\delta\cap\ch^k(M)^\perp}\!\!\!\!\!\!\!\!\!\!\!
{\rm e}^{-(a\:,\:c_1\Delta_ka)^2_{L^2(M)}-
\left(a\:,\:c_2\Delta_ka\right)_{L^2(M)}}
\:\DD a:=\frac{(\sqrt{-1}K(\delta))^{\frac{1}{2}{\rm rk'}(c_1\Delta_k)}}
{2\sqrt{\pi}\sqrt{{\rm det'}(c_1\Delta_k)}}\int\limits_{\gamma_s}
\left(t+\sqrt{-1}\:\frac{c_2}{c_1}t\right)^{-\frac{1}{2}{\rm rk'}(c_1\Delta_k)}
{\rm e}^{-\frac{t^2}{4}}\:\dd t\:\:.
\label{nemcsavartaltalanosgauss}
\end{equation}

\noindent We will be also assuming that the following 
``monotonicity principles'' hold true for these infinite dimensional 
formal integrals: 
\vspace{0.1in}

\noindent{\bf Monotonicity principles}. {\it If 
$\emptyset\subseteqq A,B\subseteqq L^2(S^4;\wedge^1S^4)$ are two 
``measurable'' subsets in the $L^2$ Hilbert space
of $1$-forms over the $4$-sphere satisfying 
$A\subseteqq B$ and $f: L^2(S^4;\wedge^1S^4)\rightarrow\R$ 
is a non-negative ``integrable'' function then
\[0\leqq\int\limits_Af(a)\DD a\leqq\int\limits_Bf(a)\DD a\leqq+\infty\:\:.\]
Moreover, if $f,g:L^2(S^4;\wedge^1S^4)\rightarrow\R$ are two 
``integrable'' functions satisfying $0\leqq f\leqq g$ then
\[0\leqq\int\limits_Af(a)\DD a\leqq\int\limits_Ag(a)\DD a\leqq+\infty\]
is valid.}
\vspace{0.1in}

\begin{remark}\rm 1. As we mentioned before the ``principal axis hypercube'' 
$C_\delta\subset L^2(M;\wedge^kM)$ for the Laplacian is not open in 
infinite dimensions. If nevertheless the formal integral 
(\ref{nemcsavartaltalanosgauss}) happens to attain a non-zero value 
then this would imply that infinite dimensional integration over very 
small (i.e. which do not contain any open ball) subsets might yield 
non-trivial results. 

2. The monotonicity properties of integration are
straighforward in finite dimensions however are not easily accessable in
infinite dimensions. But more surprisingly, it seems
these properties even may not hold over any $4$-manifold. For instance, as we
will see in Section \ref{three}, over the $4$-sphere the regularized dimension
of $L^2(S^4;\wedge^1S^4)$ with respect to the Laplacian is {\it positive}
(see Lemma \ref{szamolas}) hence the above monotonicity properties are
expected to hold true. However, over the flat $4$-torus for example, the
regularized dimension of $L^2(T^4;\wedge^1T^4)$ with respect to the Laplacian
is {\it negative} hence one would expect that some sort of reversed form
of the above monotonicity might work in this case.

All of these oddities of integration in infinite dimensions 
likely are connected with the conflict between $\sigma$-additivity 
and infinite dimensionlity (cf. the Appendix here).
\end{remark}


\section{The partition function about the vacuum}
\label{three}


After these preliminaries we are ready to calculate the partition function. 
Let us begin with recalling and introducing $4$ dimensional Euclidean 
non-supersymmetric ${\rm SU}(2)$ gauge theory with $\theta$ term in the usual 
way. 

Consider $\R^4$ with its standard flat Euclidean metric $\eta$. Let 
$E\cong\R^4\times\C^2$ be the unique trivial complex rank-two ${\rm SU}(2)$ 
vector bundle over $\R^4$ and take a compatible (i.e. ${\rm SU}(2)$-valued) 
connection $\nabla$ on it. Denoting by $\wedge^k\R^4\otimes{\mathfrak s}
{\mathfrak u}(2)$ the bundle of ${\mathfrak s}{\mathfrak u}(2)$-valued 
$k$-forms over $\R^4$, by the global triviality of $E$ we can globally write 
$\nabla =\dd +A$ where the gauge potential $A$ is a section of 
$\wedge^1\R^4\otimes{\mathfrak s}{\mathfrak u}(2)$ with the corresponding 
field strength $F_\nabla =\dd A+A\wedge A$ giving rise to a section of 
$\wedge^2\R^4\otimes{\mathfrak s}{\mathfrak u}(2)$. 
Moreover let $e\in\R$ and $\theta\in\R$ denote the coupling constant and the 
$\theta$-parameter of the theory respectively. The non-supersymmetric 
$4$ dimensional Euclidean ${\rm SU}(2)$ gauge theory is 
then defined by the action   
\[S(\nabla ,e,\theta ):=-\frac{1}{2e^2}\int\limits_{\R^4} {\rm tr}
(F_\nabla\wedge *F_\nabla)
+\frac{\sqrt{-1}\:\theta}{16\pi^2}\int\limits_{\R^4} {\rm tr}(F_\nabla\wedge 
F_\nabla)\:\:.\]
The $\theta$-term is a characteristic class hence its variation is
identically zero consequently the Euler--Lagrange equations 
(togehter with the Bianchi identity) of this theory are nothing but 
the usual vacuum Yang--Mills equations 
\[\left\{\begin{array}{ll}
\dd_\nabla F_\nabla =0\\
\dd^*_\nabla F_\nabla =0\:\:.
                       \end{array}\right.\]
Introducing the complex coupling constant 
\begin{equation}
\tau :=\frac{\theta}{2\pi}+\frac{4\pi}{e^2}\sqrt{-1}
\label{tau}
\end{equation}
taking its values on the upper complex half-plane $\C^+$, and 
the {\it positive} definite $L^2$ scalar product 
$(\Phi,\Psi)_{L^2(\R^4)}:=-\int_{\R^4}{\rm tr}(\Phi\wedge *\Psi)$ 
on the space of ${\mathfrak s}{\mathfrak u}(2)$-valued $2$-forms, 
with induced norm therefore satisfying $\Vert\Phi\Vert_{L^2(\R^4)}\geqq 0$, the 
action above can be re-written as 
\begin{eqnarray}
S(\nabla ,\tau )&=&
-\frac{\sqrt{-1}\pi}{2}\tau \left(\frac{1}{8\pi^2}
\Vert F_\nabla\Vert^2_{L^2(\R^4)}+
\frac{1}{8\pi^2}(F_\nabla,*F_{\nabla})_{L^2(\R^4)}\right)\nonumber\\
&&+\frac{\sqrt{-1}\pi}{2}\overline{\tau}\left(\frac{1}{8\pi^2}
\Vert F_\nabla\Vert^2_{L^2(\R^4)}-
\frac{1}{8\pi^2}(F_\nabla,*F_{\nabla})_{L^2(\R^4)}\right)
\label{hatas2}
\end{eqnarray}
since $*^2={\rm Id}_{\wedge^2\R^4}$ hence the topological 
term takes the shape $-\int_{\R^4}{\rm tr}(F_\nabla\wedge F_\nabla)=
(F_\nabla ,*F_\nabla)_{L^2(\R^4)}$ in this notation.    

The orientation and the flat Euclidean metric $\eta$ on $\R^4$ is used to 
introduce various Sobolev spaces. Let $\nabla^0$ denote the trivial flat 
connection on $E$ i.e. the unique connection which satisfies 
$F_{\nabla^0}=0$. Then define 
\[\ca (\nabla^0):=\{\mbox{$\nabla$ is an ${\rm SU}(2)$ connection on $E$}
\:\vert\:\mbox{$\nabla-\nabla^0\in L^2_1(\R^4\:;\:\wedge^1\R^4
\otimes{\mathfrak s}{\mathfrak u}(2))$}\} .\]
This is the $L^2_1$ Sobolev space of ${\rm SU}(2)$ connections on $E$ 
relative to $\nabla^0$. Notice that this is a vector space (not an
affine space) and in fact $\ca (\nabla^0)\ni
\nabla\mapsto\nabla-\nabla^0=:a\in L^2_1(\R^4\:;\:\wedge^1\R^4\otimes
{\mathfrak s}{\mathfrak u}(2))$ is a canonical isomorphism between 
$\ca (\nabla^0)$ and $L^2_1(\R^4\:;\:\wedge^1\R^4
\otimes{\mathfrak s}{\mathfrak u}(2))$. Furthermore write $\csu$ for the 
$L^2_2$ completion of 
\[\{\mbox{$\gamma$ is an ${\rm SU}(2)$ gauge transformation on $E$}
\:\vert\:\mbox{$\gamma -{\rm Id}_E\in C^\infty _0(\R^4;{\rm End}E)$,
$\gamma\in C^\infty (\R^4;{\rm Aut}E)$ a.e.}\}\]
that is, the space of compactly supported smooth ${\rm SU}(2)$ gauge 
transformations. Therefore $\gamma\in\csu$ means that 
$\Vert\gamma -{\rm Id}_E\Vert_{L^2_2(\R^4)}<+\infty$. The space 
$\ca (\nabla^0)$ is acted upon by $\csu$ as 
$\nabla\mapsto \gamma^{-1}\nabla\gamma$ in the usual way and the 
corresponding gauge equivalence class of $\nabla\in\ca (\nabla^0)$ 
is denoted by $[\nabla]$ and the orbit space $\ca (\nabla^0)/\csu$ of these 
equivalence classes with its quotient topology by $\cb ([\nabla^0])$ as usual. 
In the non-Abelian case $\cb ([\nabla^0])$ is not a linear space however at 
least locally it can be modeled on various Banach spaces as we shall see 
shortly. Also note that $\nabla\in\ca (\nabla^0)$ implies that if $\nabla =
\nabla^0+a$ then both the derivative term $\dd a$ and by the 
Sobolev multiplication theorem $L^2_1\times L^2_1\rightarrow L^2$ the 
interacting term $a\wedge a$ belong to $L^2$ therefore 
$F_\nabla\in L^2(\R^4\:;\:\wedge^2\R^4\otimes{\mathfrak s}{\mathfrak u}(2))$ 
for any $\nabla\in[\nabla]\in\cb ([\nabla^0])$.  

Having now the classical non-supersymmetric Euclidean gauge theory at our 
disposal, the {\it partition function} of the induced quantum theory is 
formally defined by the integral 
\[Z(\R^4,\tau ):=\frac{1}{{\rm Vol}\left({\csu}\right)}
\int\limits_{\nabla\in\ca (\nabla^0)}{\rm e}^{-S(\nabla,\tau )}\DD\nabla\]
or formally equivalently
\[Z(\R^4,\tau ):=
\int\limits_{[\nabla]\in\cb ([\nabla^0])} {\rm e}^{-S(\nabla,\tau )}\DD 
[\nabla]\]
where $\DD\nabla$ is the formal (probably never definable) measure on 
$\ca (\nabla^0)$ while $\DD[\nabla]$ is the induced formal measure 
(including the Faddeev--Popov determinant) on the orbit space 
$\cb ([\nabla^0])$. The ideal goal would be to calculate this integral in its 
full glory however it is an extraordinary difficult task because of the 
non-linearity of $\cb ([\nabla^0])$. Therefore we will evaluate it in 
$\cb_{\varepsilon} ([\nabla^0])$ only i.e. we 
are interested in an appropriately {\it truncated Feynman integral} 
\[Z_{\varepsilon} (\R^4,\tau):=\int\limits_{[\nabla]\in\cb_
{\varepsilon} ([\nabla^0])} {\rm e}^{-S(\nabla,\tau )}\DD [\nabla]\]
where $\cb_{\varepsilon}([\nabla^0])$ is a small open subset 
about $[\nabla^0]$ defined by 
$0\leqq\vert S(\nabla ,\tau)\vert<\frac{\vert\tau\vert}{8\pi}\varepsilon^2$ 
possessing the crucial property that, unlike the whole $\cb([\nabla^0])$, it 
is well approximated by (a quotient of) a small open ball in an appropriate 
Hilbert space. 

To make this picture more precise and in order to avoid several technical 
difficulties we make a technical interlude 
and extend the ${\rm SU}(2)$ Yang--Mills theory from $(\R^4,\eta)$ to its 
one-point conformal compactification $(S^4,g_R)$ where $g_R$ denotes the 
standard round metric on $S^4=\R^4\cup\{\infty\}$ such that it has radius 
$0<R<+\infty$. From the physical viewpoint this conformal compactification is 
justified at least classically by the conformal invariance of classical gauge 
theory defined by (\ref{hatas2}) in four dimensions. From the mathematical or 
technical viewpoint a further support is Uhlenbeck's singularity removal 
theorem \cite{uhl1} or rather its generalization \cite[Corollary 2.2]{uhl3} 
asserting that if $\nabla\in\ca (\nabla^0)$ is {\it any} connection on $\R^4$ 
(which by definition means that there exists an $L^2_1$ gauge relative to 
$\nabla^0$ implying $\Vert F_\nabla\Vert_{L^2(\R^4)}<+\infty$ as we 
mentioned above) there exists an $L^2_2$ gauge transformation around the 
asymptotic region of $\R^4$ such that the gauge transformed connection 
$\nabla'$ extends over $\R^4\cup\{\infty\}=S^4$. Therefore, from 
now on, instead of $(\R^4,\eta)$ we consider the classical Yang--Mills 
theory (\ref{hatas2}) over $(S^4,g_R)$ and treat $R$ as a technical parameter 
of the original theory; correspondingly we are interested in calculating the 
formal truncated Feynman integral $Z_{\varepsilon}(\R^4,\tau)$ by working 
over $(S^4,g_R)$. It is therefore understood that the action $S$, 
the Sobolev space $\ca(\nabla^0)$ consisting of our connections $\nabla$ and 
the various differential operators like $\dd^*$, $\Delta_k$, etc. are defined 
over the round $4$-sphere $(S^4,g_R)$ from now on. In this 
compactified setting Uhlenbeck's gauge fixing theorem \cite{uhl2} can be 
formulated as follows (cf. \cite[Proposition 2.3.13]{don-kro}): There 
exists a constant $0<\varepsilon$ such that 
if a connection $\nabla\in\ca (\nabla^0)$ on the trivial 
bundle $E\cong S^4\times\C^2$ satisfies $\Vert F_\nabla
\Vert_{L^2(S^4,g_R)}<\varepsilon$ then there exists an $L^2_2$ gauge 
transformation $\gamma$ and a constant $0<N(R)<+\infty$ such that the gauge 
transformed connection $\nabla'=\gamma^{-1}\nabla\gamma$ with corresponding 
decomposition $\nabla'=\dd +A'$ satisfies the {\it Coulomb gauge condition} 
together with an estimate
\begin{equation}
\left\{\begin{array}{ll}
\dd^*A'=0\\
\Vert A'\Vert_{L^2_1(S^4,g_R)}\leqq N(R)\Vert F_{\nabla'}
\Vert_{L^2(S^4,g_R)}
             \end{array}\right.
\label{coulomb}
\end{equation}
implying $\Vert A'\Vert_{L^2_1(S^4,g_R)}\leqq N(R)\varepsilon$ in Coulomb gauge.

Now we are in a position to define the truncated partition function more 
carefully. Take a constant $0<\varepsilon <\sqrt{8}\pi$ and consider those 
connections which satisfy $\Vert F_\nabla\Vert_{L^2(S^4,g_R)}<\varepsilon$. 
By conformal invariance of the norm this is equivalent to consider those 
connections over the original space which satisfy 
$\Vert F_\nabla\Vert_{L^2(\R^4)}<\varepsilon$. The action takes a more clear 
shape in the compactified setting as follows. Regarding its topological 
term $\frac{1}{8\pi^2}\int_{S^4}{\rm tr}(F_\nabla\wedge F_\nabla)$ we know 
that it is proportional to the second Chern 
number of the extended ${\rm SU}(2)$ bundle over $S^4$ hence it assumes 
integer values only; however by the Cauchy--Schwarz inequality 
$0\leqq\left\vert(F_\nabla\:,\:*F_\nabla)_{L^2(S^4,g_R)}\right\vert\leqq 
\Vert F_\nabla\Vert^2_{L^2(S^4,g_R)}<\varepsilon^2<8\pi^2$ 
the $\theta$-term simply vanishes over $S^4$ in the small energy regime. 
This also implies that the connections we are interested 
in are realized in the extended gauge theory on the trivial bundle 
$E\cong S^4\times\C^2$ alone and if $\varepsilon$ is small enough then 
Uhlenbeck's gauge fixing theorem applies. Consequently the action 
(\ref{hatas2}) about $\nabla^0$ reduces to  
\[S(\nabla ,\tau)=S(\nabla^0+a,\tau)=\frac{{\rm Im}\tau}{8\pi}
\Vert F_{\nabla^0+a}\Vert^2_{L^2(S^4,g_R)}=\frac{{\rm Im}\tau}{8\pi}
\Vert\dd a+a\wedge a\Vert^2_{L^2(S^4,g_R)}\]
which also shows by conformal invariance of the action that 
$[\nabla]\in\cb_\varepsilon([\nabla^0])$. The key technical observation now is 
\cite[Proposition 4.2.9]{don-kro} saying that for a sufficiently small 
$\varepsilon$ there exists an $\eta>0$ such that $\cb_\varepsilon 
([\nabla^0])$ is homeomorphic to $(B_\eta (\nabla^0)\cap{\rm 
ker}\:\dd^*)/G_0$ with $B_\eta (\nabla^0)\subset\ca (\nabla^0)$ being a 
small open ball and $G_0\cong{\rm SU}(2)$ the gauge isotropy subgroup of the 
flat hence reducible connection $\nabla^0$. Hence put 
\begin{equation}
\ca_{\varepsilon,N(R)}(\nabla^0):=\left\{a\in L^2_1(S^4;\wedge^1S^4\otimes
{\mathfrak s}{\mathfrak u}(2))\:\left\vert\:\Vert a\Vert_{L^2_1(S^4,g_R)}
<\min (\eta, N(R)\varepsilon )\right.\right\}
\label{vakuumgolyo}
\end{equation}
where $\varepsilon, N(R)$ are the same constants over $(S^4,g_R)$ 
as in (\ref{coulomb}). By the aid of the homeomorphism 
\[\cb_\varepsilon ([\nabla^0])\cong\frac{\ca_{\varepsilon,N(R)}(\nabla^0)
\cap{\rm ker}\:\dd^*}{G_0}\] 
we {\it suppose} that the ``measure'' $\DD[\nabla]$ arises from 
a $G_0$-invariant ``measure'' on $\ca_{\varepsilon,N(R)}
(\nabla^0)\cap{\rm ker}\:\dd^*$ what we denote $\DD_Ra$. The main advantage of 
this non-linear isomorphism is 
that it locally ``straightens'' the gauge orbits hence its effect is 
analogous to passing from a general curved coordinate system to the standard 
Descartes one. Consequently the {\it Faddeev--Popov determinant} is locally 
transformed away i.e. gives only a constant multiplyer (cf. 
Footnote \ref{harom}). Moreover the {\it Gribov ambiguity} problem does not 
cause any headache here too, for this local quotient contains nearby gauge 
orbits precisely once only. Our truncated Feynman integral awaiting 
for computation is then defined more carefully as 
\begin{equation} 
Z_{\varepsilon}(\R^4,\tau ):=\int\limits_{\ca_{\varepsilon,N(R)} 
(\nabla^0)\cap{\rm ker}\:\dd^*}{\rm e}^{-\frac{{\rm Im}\tau}{8\pi}
\Vert\dd a+a\wedge a\Vert^2_{L^2(S^4,g_R)}}\frac{\DD_Ra}{{\rm Vol}(G_0)}
\label{particiofv1}
\end{equation}
having the following properties. In this formal integral the 
integration domain $\ca_{\varepsilon,N(R)}(\nabla^0)\cap{\rm ker}\:\dd^*$ 
is a small open ball of radius $\min (\eta,N(R)\varepsilon)$ in the (by the 
compactness of $S^4$) closed hence Hilbert subspace 
${\rm ker}\:\dd^*\subset L^2_1(S^4;\wedge^1S^4
\otimes{\mathfrak s}{\mathfrak u}(2))$; consequently the size of this ball 
depends on the radius $R$ through the Uhlenbeck constant $N(R)$ in 
(\ref{coulomb}). Moreover, in this formal integral the hypothetical 
integration ``measure'' 
$\DD_Ra$ may in principle depend on the radius $R$ of $S^4$ too. Consequently, 
in spite of the conformal invariance of the integrand 
${\rm e}^{-\frac{{\rm Im}\tau}{8\pi}\Vert\dd a+a\wedge a
\Vert^2_{L^2(S^4,g_R)}}$, the formal integral itself may fail to be 
conformally invariant (cf. Lemma \ref{eredmeny}). For notational simplicity 
we shall hide both the numerical factor $0<\frac{1}{{\rm Vol}(G_0)}<+\infty$ 
and the $R$-dependence and denote $\frac{\DD_Ra}{{\rm Vol}(G_0)}$ simply 
as $\DD a$ from now on. 

Let us work out a two-sided estimate for the action appering in 
(\ref{particiofv1}) but along a perhaps resized integration domain as follows.  
\begin{lemma} 
For every fixed finite value $0<{\rm Im}\tau<+\infty$ of the imaginary part of 
complex coupling constant (\ref{tau}) there exists a sufficiently small but 
yet finite value of the vicinity parameter $\varepsilon$ such that 
(\ref{coulomb}) is applicable and there exist constants $0<N,c<+\infty$ where   
\[N:=\lim\limits_{R\rightarrow 0}\left(\inf\big\{N(R)\:\big\vert\:
\Vert a\Vert_{L^2_1(S^4,g_R)}\leqq N(R)
\Vert\dd a+a\wedge a\Vert_{L^2(S^4,g_R)}\:,\:
a\in\ca_{\varepsilon, N(R)}(\nabla^0)\cap{\rm ker}\dd^*\big\}\right)\]
such that for every $a\in \ca_{\varepsilon, N}(\nabla^0)\cap{\rm ker}\:\dd^*$ 
in the correspondingly resized ball a two-sided estimate
\begin{equation}
\frac{2\:{\rm Im\tau}}{\pi N^4}\Vert\dd a\Vert^4_{L^2(S^4)}\leqq
\Vert\dd a+a\wedge a\Vert^2_{L^2(S^4)}\leqq 2\Vert\dd a\Vert^2_{L^2(S^4)}+
2c^2\Vert\dd a\Vert^4_{L^2(S^4)}
\label{uhlenbeck}
\end{equation}
holds in Coulomb gauge. 

Note that all norms in this inequality are conformally invariant. 
Accordingly, both $0<N,c<+\infty$ are conformally invariant and 
$1\leqq N$ such that $N\rightarrow 1$ as $\varepsilon\rightarrow 0$. 
\label{kettosegyenlotlenseglemma}
\end{lemma}

\begin{proof} We begin with the estimate from below in 
(\ref{uhlenbeck}) which, as often happens, is much more difficult than 
obtaining an estimate from above. 

Assume that $\varepsilon$ is small enough hence (\ref{coulomb}) is 
applicable; it readily follows that working over the unit 
sphere $(S^4,g_1)$ we have $\Vert a\Vert_{L^2_1(S^4,g_1)}=
\Vert a\Vert_{L^2(S^4,g_1)}+\Vert\dd a\Vert_{L^2(S^4,g_1)}\leqq N(1) 
\Vert\dd a+a\wedge a\Vert_{L^2(S^4,g_1)}$. Observe that in this inequality both 
$\Vert\dd a\Vert_{L^2(S^4,g_1)}$ and $\Vert\dd a+a\wedge a\Vert_{L^2(S^4,g_1)}$ 
are conformally invariant, thus we shall denote them respectively as 
$\Vert\dd a\Vert_{L^2(S^4)}$ and $\Vert\dd a+a\wedge a\Vert_{L^2(S^4)}$ 
from now on, while $\Vert a\Vert_{L^2(S^4,g_1)}$ is not. More 
precisely, if we pass to $(S^4,g_R)$ then the latter norm scales as 
$\Vert a\Vert_{L^2(S^4,g_R)}=R\Vert a\Vert_{L^2(S^4,g_1)}$. 
Consequently defining $N$ by taking the limit $R\rightarrow 0$ as above 
we obtain an inequality  
\begin{equation}
\Vert\dd a\Vert_{L^2(S^4)}\leqq N\Vert \dd a+a\wedge a\Vert_{L^2(S^4)}
\label{egyenlotlenseg1}
\end{equation}
over the appropriately resized ball
$\ca_{\varepsilon, N}(\nabla^0)\cap{\rm ker}\:\dd^*$ 
having the following properties. This $N$ is optimal and universal in the sense 
that it is the smallest available constant (at least in the Uhlenbeck setting) 
hence satisfies $N\leqq N(R)$ for any Uhlenbeck constant from (\ref{coulomb}) 
over $(S^4,g_R)$ moreover $N$ is conformally invariant. 

Taking Abelian $1$-forms i.e. $a\in\ca_{\varepsilon, N}(\nabla^0)\cap
\ker\dd^*$ which satisfy $a\wedge a=0$ a.e. then (\ref{egyenlotlenseg1}) 
shows that $\Vert\dd a\Vert_{L^2(S^4)}\leqq N\Vert\dd a \Vert_{L^2(S^4)}$ 
moreover knowing that by the Coulomb gauge condition $\dd a=0$ if and only if 
$a=0$ a.e. on the one hand $1\leqq N$. In the generic non-Abelian case 
$\Vert a\wedge a\Vert_{L^2(S^4)}$ is bounded by $\Vert a\Vert^2_{L^4(S^4)}$; 
but $\Vert a\Vert_{L^4(S^4)}\!\leqq\!c_1\Vert a\Vert_{L^2_1(S^4)}$ by the 
Sobolev embedding $L^2_1\subset L^4$ which is sharp in $4$ dimensions; moreover 
elliptic regularity for $\dd +\dd^*$ gives $\Vert a\Vert_{L^2_1(S^4)}\leqq 
c_2\Vert (\dd +\dd^*)a\Vert_{L^2(S^4)}+c_3\Vert a\Vert_{L^2(S^4)}=
c_2\Vert\dd a\Vert_{L^2(S^4)}$ since $\dd^*a=0$ by the Coulomb
gauge condition and we can put $c_3=0$ because $H^1(S^4)=0$ consequently 
${\rm ker}(\dd +\dd^*)={\rm ker}\:\Delta_1=\{0\}$. Combining these 
and introducing $c:=(c_1c_2)^2>0$ we get
\begin{equation}
\Vert\dd a\Vert_{L^2(S^4)}\leqq\Vert\dd a+a\wedge a\Vert_{L^2(S^4)}+
\Vert a\wedge a\Vert_{L^2(S^4)}\leqq \Vert\dd a+a\wedge a\Vert_{L^2(S^4)}+
c\Vert\dd a\Vert^2_{L^2(S^4)}\:\:.
\label{egyenlotlenseg2}
\end{equation}
Regarding the constant $c$ note that it says 
$\Vert a\Vert_{L^4(S^4)}\leqq\sqrt{c}\:\Vert\dd a\Vert_{L^2(S^4)}$ and both 
norms here are conformally invariant hence we can assume that $c$ is 
conformally invariant as well. Proceeding further, 
by the aid of (\ref{coulomb}) take any $N(R)\geqq 1$ satisfying 
$\Vert\dd a\Vert_{L^2(S^4)}\leqq\Vert a\Vert_{L^2_1(S^4,g_R)}\leqq 
N(R)\Vert\dd a+a\wedge a\Vert_{L^2(S^4)}$ over $(S^4,g_R)$.
Adding the two estimates for $\Vert\dd a\Vert_{L^2(S^4)}$ provided by 
(\ref{egyenlotlenseg2}) and this last inequality we obtain 
$\Vert\dd a\Vert_{L^2(S^4)}(2-c\Vert\dd
a\Vert_{L^2(S^4)})\leqq (N(R)+1)\Vert\dd a+a\wedge a\Vert_{L^2(S^4)}$. 
Moreover we have $\Vert\dd a\Vert_{L^2(S^4)}<N(R)\varepsilon$ in 
(\ref{vakuumgolyo}) thus $\Vert\dd a\Vert_{L^2(S^4)}
\left(2-cN(R)\varepsilon\right)
\leqq (N(R)+1)\Vert \dd a+a\wedge a\Vert_{L^2(S^4)}$. Provided 
$\varepsilon$ is small enough compared with the initial value of $N(R)$, 
more precisely if $\varepsilon <\frac{2}{cN(R)}$ then we can replace 
$N(R)$ with $\frac{N(R)+1}{2-cN(R)\varepsilon}$ and iterate this process; 
the general theory of iteration guarantees that $N(R)$ will converge to the 
lower fixed point $N_*(R)=\frac{1}{2c\varepsilon}\big(1-\sqrt{1-4c\varepsilon}
\:\big)$ of this iteration. Since $N$ from (\ref{egyenlotlenseg1}) 
is the optimal constant we have on the other hand $N\leqq N_*(R)$ consequently 
\[1\leqq N\leqq N_*(R)=1+\frac{1}{4}c\varepsilon+\dots\]
demonstrating that $N\rightarrow 1$ as $\varepsilon\rightarrow 0$. Assume that 
$0<\varepsilon<\sqrt{\frac{\pi}{2\:{\rm Im}\tau}}$ then 
$\Vert\dd a\Vert_{L^2(S^4)}<N\varepsilon <N\sqrt{\frac{\pi}
{2\:{\rm Im}\tau}}$ within the ball 
$\ca_{\varepsilon, N}(\nabla^0)\cap\ker\dd^*$ consequently multiplying the 
inequality (\ref{egyenlotlenseg1}) by $\Vert\dd a\Vert_{L^2(S^4)}$ we get
\[\Vert\dd a\Vert^2_{L^2(S^4)}\leqq 
N\Vert\dd a+a\wedge a\Vert_{L^2(S^4)}\Vert\dd a\Vert_{L^2(S^4)}\leqq
\sqrt{\frac{\pi}{2\:{\rm Im}\tau}}\:N^2\Vert\dd a+a\wedge a\Vert_{L^2(S^4)}\] 
hence squaring it we come up with the estimate from below in (\ref{uhlenbeck}). 

The estimate from above is simpler. 
We start with $\Vert\dd a+a\wedge a\Vert^2_{L^2(S^4)}\leqq 
2\Vert\dd a\Vert^2_{L^2(S^4)}+2\Vert a\wedge 
a\Vert^2_{L^2(S^4)}$ and then repeat the steps towards 
(\ref{egyenlotlenseg2}) to end up with 
\[\Vert\dd a+a\wedge a\Vert^2_{L^2(S^4)}\leqq 2\Vert\dd a\Vert^2_{L^2(S^4)}+
2c^2\Vert\dd a\Vert^4_{L^2(S^4)}\]
where $c$ is the conformally invariant constant used so far. Letting for 
instance 
\[\varepsilon:=\frac{1}{2}
\min\left(\sqrt{8}\pi\:,\:\mbox{the original Uhlenbeck condition 
in (\ref{coulomb})}\:,\:\frac{2}{cN(R)}\:,\:\sqrt{\frac{\pi}{2\:{\rm Im}\tau}}
\:\:\right)\]
and then putting together the last two estimates we obtain 
the desired two-sided inequality. 
\end{proof}

\noindent Let us proceed further by multiplying each term in (\ref{uhlenbeck}) 
with $-\frac{{\rm Im}\tau}{8\pi}<0$ and then exponentiating: 
\[{\rm e}^{-\left(\frac{{\rm Im}\tau}{2\pi N^2}\right)^2\Vert\dd 
a\Vert^4_{L^2(S^4)}}\geqq 
{\rm e}^{-\frac{{\rm Im}\tau}{8\pi}\Vert\dd a+a\wedge a\Vert^2_{L^2(S^4)}}
\geqq {\rm e}^{-\frac{{\rm Im}\tau}{4\pi}\Vert\dd a\Vert^2_{L^2(S^4)}
-\frac{{\rm Im}\tau\:c^2}{4\pi}\Vert\dd a\Vert^4_{L^2(S^4)}}\]
or equivalently, using $\dd^*a=0$ again
\begin{equation}
{\rm e}^{-\left(a\:,\:\frac{{\rm Im}\tau}{2\pi N^2}\Delta_1a
\right)^2_{L^2(S^4)}}\geqq 
{\rm e}^{-\frac{{\rm Im}\tau}{8\pi}\Vert\dd a+a\wedge a\Vert^2_{L^2(S^4)}}
\geqq {\rm e}^{-\left(a\:,\:\sqrt{\frac{{\rm Im}\tau}{4\pi}}c\:\Delta_1a
\right)^2_{L^2(S^4)}-\left(a\:,\:\frac{{\rm Im}\tau}{4\pi}\Delta_1a\right)_
{L^2(S^4)}}\:\:.
\label{kulcsegyenlotlenseg}
\end{equation}
Having obtained these rigorous estimates consider the vicinity 
of the vacuum in Coulomb gauge i.e. the small ball about
the flat connection $\ca_{\varepsilon,N}(\nabla^0)\cap{\rm ker}\:\dd^*$ 
as in (\ref{vakuumgolyo}) however such that $N(R)$ in its radius has been 
replaced with the universal $N$ from (\ref{uhlenbeck}). Take the Laplacian 
$\Delta_1$ and the corresponding $C_\delta\subset L^2_1(S^4;\wedge^1S^4)$ 
introduced as its finite dimensional analogue (\ref{hiperkocka}). If
$0<\lambda_{{\rm min}}<+\infty$ is the smallest eigenvalue of
$\Delta_1$ then picking any $0<\delta <\sqrt{\lambda_{{\rm min}}}
\:\min(\eta,N\varepsilon)$ we know by (\ref{vakuumgolyo}) that
$\ca_{\varepsilon,N}(\nabla^0)\supset
C_\delta$ yielding two inclusions ${\rm ker}\:\dd^*\supset
\ca_{\varepsilon,N}(\nabla^0)\cap
{\rm ker}\:\dd^*\supset C_\delta\cap{\rm ker}\:\dd^*$ 
for these subsets in $L^2_1(S^4;\wedge^1S^4)$. Now let us formally 
integrate the left term of (\ref{kulcsegyenlotlenseg}) over ${\rm ker}\:\dd^*$, 
the middle term of (\ref{kulcsegyenlotlenseg}) over 
$\ca_{\varepsilon,N}(\nabla^0)\cap {\rm ker}\:\dd^*$ and finally the right 
term of (\ref{kulcsegyenlotlenseg}) over $C_\delta\cap{\rm ker}\:\dd^*$. 
Referring at this step to our {\bf Monotonicity principles} this procedure 
obeys the ordering in (\ref{kulcsegyenlotlenseg}) thus formally 
\[\begin{array}{ll}
\int\limits_{{\rm ker}\:\dd^*}\!\!\!\!\!\!
{\rm e}^{-\left(a,\frac{{\rm Im}\tau}{2\pi N^2}\Delta_1a
\right)^2_{\!\!L^2(S^4)}}\DD a
\geqq\!\!\int\limits_{\ca_{\varepsilon,N}(\nabla^0)\cap {\rm ker}\:\dd^*}
\!\!\!\!\!\!\!\!\!\!\!\!\!\!\!\!\!\!\!\!\!\!
{\rm e}^{-\frac{{\rm Im}\tau}{8\pi}\Vert\dd a+a\wedge a\Vert^2_{L^2(S^4)}}
\DD a\geqq\int\limits_{C_\delta\cap{\rm ker}\:\dd^*}\!\!\!\!\!\!\!\!\!\!\!
{\rm e}^{-\left(a,\sqrt{\frac{{\rm Im}\tau}{4\pi}}c\:\Delta_1a
\right)^2_{\!\!L^2(S^4)}\!\!\!\!\!\!\!\!-\left(a,\:\frac{{\rm
Im}\tau}{4\pi}\Delta_1a\right)_{L^2(S^4)}}\DD a
\end{array}\]
continues to hold. The time has come to apply our formal integral expressions 
from Section \ref{two}.

\begin{definition} {\rm (cf. \cite[Definition 3.1]{ete-nag})} 
Taking into account that $H^1(S^4)=\{0\}$ and 
$\dim_\R{\mathfrak s}{\mathfrak u}(2)=3$ substituting 
$c:=\frac{{\rm Im}\tau}{2\pi N^2}$ in (\ref{nemcsavartgauss}) we define 
a non-truncated quartic integral as 
\[\int\limits_{{\rm ker}\:\dd^*}
{\rm e}^{-\left(a\:,\:\frac{{\rm Im}\tau}{2\pi N^2}
\Delta_1a\right)^2_{L^2(S^4)}}\:\DD a:=
\left(\frac{(\sqrt{-1}\pi)^{\frac{1}{2}{\rm rk'}
\left(\frac{{\rm Im}\tau}{2\pi N^2}
\Delta_1\vert_{{\rm ker}\:\dd^*}\right)}}{\sqrt{{\rm det'}\left(
\frac{{\rm Im}\tau}{2\pi N^2}\Delta_1\vert_{{\rm ker}\:\dd^*}\right)}}
\right)^{3}\:\:\frac{1}{2\sqrt{\pi}}
\int\limits_{\gamma_s}t^{-\frac{3}{2}{\rm rk'}\left(
\frac{{\rm Im}\tau}{2\pi N^2}
\Delta_1\vert_{{\rm ker}\:\dd^*}\right)}{\rm e}^{-\frac{t^2}{4}}\dd t\:\:.\]
Likewise, substituing $c_2:=\frac{{\rm Im}\tau}{4\pi}$ and $c_1:=
\sqrt{\frac{{\rm Im}\tau}{4\pi}}\:c$ in (\ref{nemcsavartaltalanosgauss}) 
we define a truncated quartic integral
\begin{eqnarray}
\int\limits_{C_\delta\cap{\rm ker}\:\dd^*}\!\!\!\!\!\!\!\!
{\rm e}^{-\left(a\:,\:\sqrt{\frac{{\rm Im}\tau}{4\pi}}c\:\Delta_1a
\right)^2_{L^2(S^4)}-\left(a\:,\:\frac{{\rm Im}\tau}{4\pi}\Delta_1a\right)_
{L^2(S^4)}}\DD a&\!\!\!\!\!:=\!\!\!\!\!&
\left(\frac{(\sqrt{-1}K(\delta))^{\frac{1}{2}
{\rm rk'}\left(\frac{{\rm Im}\tau}{4\pi}\Delta_1\vert_{{\rm ker}\:
\dd^*}\right)}}{\sqrt{{\rm det'}\left
(\frac{{\rm Im}\tau}{4\pi}
\Delta_1\vert_{{\rm ker}\:\dd^*}\right)}}\right)^{3}\times\nonumber\\
&&\frac{1}{2\sqrt{\pi}}\int\limits_{\gamma_s}
\left(t+\sqrt{-1}\:c\sqrt{\frac{4\pi}{{\rm Im}\tau}}
\right)^{-\frac{3}{2}{\rm rk'}\left(\frac{{\rm Im}\tau}{4\pi}\Delta_1
\vert_{{\rm ker}\:\dd^*}\right)}\!\!\!{\rm e}^{-\frac{t^2}{4}}\dd t\nonumber
\end{eqnarray}
\label{definiciok}
where the common contour $\gamma_s$ is to
be specified such that to meet all demands from avoiding possible poles and 
branch cuttings in both integrals.
\end{definition}

\noindent A familiar way to make sense of ${\rm rk'}$ and
${\rm det'}$ in Definition \ref{definiciok} i.e. to regularize the dimension
and the functional determinant in infinite dimensions is an application of 
$\zeta$-function regularization. 

\begin{lemma}{\rm (cf. \cite[Lemma 3.1]{ete-nag})} 
Using $\zeta$-function regularization to define ${\rm rk'}$ 
and ${\rm det'}$ and then heat kernel techniques to calculate the 
zero values of various resulting $\zeta$-functions over $(S^4,g_R)$ we obtain  
from its definition above that the non-truncated quartic integral looks like 
\[\int\limits_{{\rm ker}\:\dd^*}
{\rm e}^{-\left(a\:,\:\frac{{\rm Im}\tau}{2\pi N^2}\Delta_1a
\right)^2_{L^2(S^4)}}\:\DD a =
\left(\frac{2\sqrt{-1}\:\pi^2N^2}{{\rm Im}\tau}
\right)^{\frac{11}{20}}
{\rm e}^{\frac{3}{2}\zeta'_{\Delta_1}(0)-3\zeta'_{\Delta_0}(0)}
\frac{1}{2\sqrt{\pi}}\int\limits_{-\infty}^{+\infty}
t^{-\frac{11}{20}}{\rm e}^{-\frac{t^2}{4}}\dd t\]
over $(S^4,g_R)$. Likewise, 
\begin{eqnarray}
\int\limits_{C_\delta\cap{\rm ker}\:\dd^*}\!\!\!\!\!\!\!\!\!
{\rm e}^{-\left(a\:,\:\sqrt{\frac{{\rm Im}\tau}{4\pi}}c\:\Delta_1a
\right)^2_{L^2(S^4)}-\left(a\:,\:\frac{{\rm Im}\tau}{4\pi}\Delta_1a\right)_
{L^2(S^4)}}\DD a&=&
\left(\frac{4\sqrt{-1}\pi K(\delta)}{{\rm Im}\tau}\right)^\frac{11}{20}\:
{\rm e}^{\frac{3}{2}\zeta'_{\Delta_1}(0)-3\zeta'_{\Delta_0}(0)}\times\nonumber\\
&&\frac{1}{2\sqrt{\pi}}\int\limits_{-\infty}^{+\infty}
\left(t+\sqrt{-1}\:c\sqrt{\frac{4\pi}{{\rm Im}\tau}}\right)^{-\frac{11}{20}}
\!\!\!\!\!{\rm e}^{-\frac{t^2}{4}}\:\dd t\nonumber
\end{eqnarray}
\label{szamolas}   
is the shape of the truncated quartic integral over $(S^4,g_R)$. 

Taking into account that the exponent in the complex 
integrals satisfies $-1<-\frac{11}{20}<0$ we know that there are no poles 
and there is a single branch cutting connecting $0$ with 
$-\sqrt{-1}\:\infty$ along the non-positive imaginary axis in the first 
integral while connecting $-\sqrt{-1}\:c\sqrt{\frac{4\pi}{{\rm Im}\tau}}$ with 
$-\sqrt{-1}\:\infty$ along the negative 
imaginary axis in the second integral. Therefore we can simply put 
$\gamma_s:=\R$ in both integrals.
\end{lemma}

\begin{remark}\rm Before embarking upon the proof we note that 
the particular value $-1<-\frac{11}{20}<0$ of the exponents in 
these integral expressions is not important because it is just the 
consequence of one of the possible (namely $\zeta$-function combined with 
heat kernel) regularization procedures carried over one of the possible 
(namely $(S^4,g_R)$ i.e. the one-point conformal) compactifications of 
$(\R^4,\eta)$. Only its sign, namely that it is negative, bears relevance. 
Indeed, this exponent does not have to always 
assume a negative value because of some {\it a priori} reason. For example in 
the case of the flat torus $T^4$ the corresponding exponent turns out to 
be $0<\frac{9}{2}<5$ leading to a completely different situation; e.g. 
the {\bf Monotonicity principles} break down due to the opposite 
scaling of the integrals. These oddities are related with lacking a good 
measure in infinite dimensions, see the Appendix. 
\end{remark} 

\begin{proof} Since the spectrum of the Laplacian over a
compact Riemannian manifold $(M,g)$ is non-negative real and discrete, one sets
\[\zeta_{\Delta_k}(s):=\sum\limits_{\lambda\in {\rm Spec}
\:\Delta_k\setminus\{0\}}\lambda^{-s},\:\:\:\:\:\mbox{with $s\in\C$ and
${\rm Re}\:s>0$ sufficiently large}\]
and observes that this function can be meromorphically continued over the
whole complex plane (cf. e.g. \cite[Theroem 5.2]{ros}) having no pole at
$s=0\in\C$. A formal calculation then convinces us that the regularized
rank and the determinant of the Laplacian should be 
${\rm rk'}\:\Delta_k:=\zeta_{\Delta_k}(0)$ and ${\rm det'}\Delta_k
:={\rm e}^{-\zeta '_{\Delta_k}(0)}$ yielding ${\rm rk'}(c\Delta_k)=
\zeta_{\Delta_k}(0)$ and ${\rm det'}(c\Delta_k)=
c^{\zeta_{\Delta_k}(0)}{\rm e}^{-\zeta '_{\Delta_k}(0)}$. 

Because of the Coulomb gauge condition we have to calculate 
restrictions of these $\zeta$-functions over the round $4$-sphere 
$(S^4,g_R)$. Since $H^1(S^4)=\{0\}$ hence $\Delta_1$ has trivial kernel, 
the Hodge decomposition theorem says that 
$L^2(S^4;\wedge^1S^4)\cong{\rm im}\:\dd_0\oplus{\rm im}\:\dd^*_2$. Moreover 
${\rm im}\:\dd_0\cap{\rm ker}\:\dd^*_1 =\{ 0\}$ and 
${\rm im}\:\dd^*_2\subseteqq{\rm ker}\:\dd^*_1$ hence 
\[L^2(S^4;\wedge^1S^4)\cong {\rm im}\:\dd_0\oplus{\rm ker}\:\dd^*_1\:\:.\]
Applying this decomposition we can
write any element $a\in L^2(S^4;\wedge^1S^4)$ uniquely as 
$a=\dd_0f +\alpha$ with $ f\in L^2_1(S^4;\wedge^0S^4)$ a function and 
$\alpha\in L^2(S^4;\wedge^1S^4)$ satisfying $\dd^*_1\alpha =0$. A simple 
calculation ensures us that
\[\left( a\:,\:\Delta_1a\right)_{L^2(S^4)}=\left(
\dd_0f +\alpha\:,\:\Delta_1(\dd_0f +\alpha)\right)_{L^2(S^4)}=
\left( f\:,\:\Delta^2_0f\right)_{L^2(S^4)}+
\left(\alpha\:,\:\Delta_1\alpha\right)_{L^2(S^4)}\]
where $\Delta^2_0$ is the square of the scalar Laplacian on $(S^4,g_R)$. 
Taking into account these decompositions then we obtain that 
${\rm Spec}\:\Delta_1={\rm Spec}\:\Delta^2_0\sqcup {\rm Spec}\:
\Delta_1\vert_{{\rm ker}\:\dd^*_1}$. This decomposition together
with the proof of \cite[Theorem 5.2]{ros} ensures us that 
$\zeta_{\Delta_1}=\zeta_{\Delta^2_0}+\zeta_{\Delta_1\vert{\rm ker}
\:\dd^*_1}$ consequently $\zeta_{\Delta_1\vert{\rm ker}
\:\dd^*}=\zeta_{\Delta_1}-\zeta_{\Delta^2_0}$. 
Moreover $\zeta_{\Delta^2_0}(s)=\zeta_{\Delta_0}(2s)$ hence 
${\rm rk'}(c\Delta^2_0)=\zeta_{\Delta_0}(0)$ and 
$\det '(c\Delta^2_0)=c^{\zeta_{\Delta_0}(0)}
{\rm e}^{-2\zeta '_{\Delta_0}(0)}$. Therefore in the case of the first 
integral of Definition \ref{definiciok} putting 
$c:=\frac{{\rm Im}\tau}{2\pi N^2}$ we find  
\[\left\{\begin{array}{ll}
{\rm rk'}\left(\frac{{\rm Im}\tau}{2\pi N^2}\Delta_1\vert_{{\rm ker}\:\dd^*}
\right)&=\zeta_{\Delta_1}(0)-\zeta_{\Delta_0}(0)\\
{\rm det'}\left(\frac{{\rm Im}\tau}{2\pi N^2}\Delta_1\vert_{{\rm ker}\:\dd^*}
\right))&=\left(\frac{{\rm Im}\tau}{2\pi N^2}\right)^{\zeta_{\Delta_1}(0)-
\zeta_{\Delta_0}(0)}{\rm e}^{-\zeta'_{\Delta_1}(0)+2\zeta'_{\Delta_0}(0)}\:\:.
\end{array}\right.\] 

We can easily calculate at least $\zeta_{\Delta_1}(0)-\zeta_{\Delta_0}(0)$ 
explicitly applying standard heat kernel techniques. Over a compact 
$4$-manifold $(M,g)$ without boundary it is well-known \cite[Theorem 5.2]{ros} 
that
\[\zeta_{\Delta_k}(0)=-\dim_\R{\rm ker}\Delta_k+
\frac{1}{16\pi^2}\int\limits_M{\rm tr}(u^4_k)\dd V_g\]
where the sections $u^p_k\in C^\infty (M;{\rm End}(\wedge^kM))$ with
$p=0,1,\dots$ appear \cite [Chapter 3]{ros} in the coefficients of the
short time asymptotic expansion of the heat kernel for the $k$-Laplacian
\[\sum\limits_{\lambda\in{\rm Spec}\Delta_k\setminus\{0\}}
{\rm e}^{-\lambda t}\:\sim\:\frac{1}{(4\pi t)^2}\sum\limits_{p=0}^{+\infty}
\left(\:\:\int\limits_M{\rm tr}(u^p_k)\dd V_g\right)
t^{\frac{p}{2}}\:\:\:\:\:\mbox{as $t\rightarrow 0$}\:\:.\]
These functions are expressible with the curvature of $(M,g)$ and one
can demonstrate \cite[p. 340]{gil} that
\[\left\{\begin{array}{ll}
u^4_0 &=\frac{1}{360}\left( 2\vert {\rm Riem}\vert_g^2-
2\vert {\rm Ric}\vert_g^2+5\:{\rm Scal}^2\right)\\
        &\\
{\rm tr}(u^4_1) &=\frac{1}{360}\left( -22\vert {\rm Riem}\vert_g^2+172
\vert {\rm Ric}\vert_g^2-40\:{\rm Scal}^2\right)
\end{array}\right.\]
yielding together with $\dim_\R{\rm ker}\:\Delta_0=1$ and 
$\dim_\R{\rm ker}\:\Delta_1=0$ over $(S^4,g_R)$ that 
\[\zeta_{\Delta_1}(0)-\zeta_{\Delta_0}(0)=
1-\frac{1}{\pi^2}\int\limits_{S^4}\left(\frac{1}{120} 
\vert{\rm Riem}\vert_g^2 -\frac{87}{2880}\vert{\rm Ric}\vert_g^2+\frac{1}{128}
{\rm Scal}^2\right)\dd V_R\:\:.\]
In addition we recall over $(S^4,g_R)$ the classical expressions 
\[\left\{\begin{array}{ll}
\vert {\rm Riem}\vert_{g_R}^2&=2\vert {\rm Ric}\vert_{g_R}^2-\frac{1}{3}\:{\rm 
Scal}^2\\
\vert{\rm Ric}\vert_{g_R}^2&=\frac{1}{4}{\rm Scal}^2\\
{\rm Scal}&=\frac{12}{R^2}
         \end{array}\right.\]    
and plug them into the integral and also perform  
$\int_{S^4}\dd V_R=\frac{8\pi^2}{3}R^4$. We come up with 
\[\frac{3}{2}\left(\zeta_{\Delta_1}(0)-\zeta_{\Delta_0}(0)\right)=
\frac{3}{2}\left(1-\frac{1}{\pi^2}\left(\frac{1}{120}
\left(\frac{1}{2}-\frac{1}{3}\right)
-\frac{87}{2880}\cdot\frac{1}{4}+\frac{1}{128}\right)\frac{144}{R^4}\cdot
\frac{8\pi^2}{3}R^4\right)=\frac{11}{20}\]
and find in particular that $\frac{3}{2}
\left(\zeta_{\Delta_1}(0)-\zeta_{\Delta_0}(0)\right)$ is independent of $R$ 
offering a sort of justification for using the conformal compactification 
$(S^4,g_R)$ in place of the original space $(\R^4,\eta)$. Inserting all of 
these formulata into the right hand side of the first integral of 
Definition \ref{definiciok} we obtain the first expression of the 
lemma.\footnote{The Faddeev--Popov determinant is therefore formally equal 
to ${\rm e}^{3\zeta'_{\Delta_0}(0)}$ hence is indeed constant in this
picture.\label{harom}} Repeating the same with the truncated quartic integral, 
the corresponding result also follows. 

The only remaining thing is to specify the common contour in the two 
complex integrals. Since $-1<-\frac{11}{20}<0$ there are 
no poles however branch cuttings required in these complex integrals as 
described hence for simplicity $\gamma_s$ can be taken to be the real line 
everywhere.
\end{proof}

\noindent By Lemma \ref{szamolas} and (\ref{particiofv1}) we
eventually arrive at the two-sided estimate
\begin{equation}
\begin{array}{cc}
\left(\frac{2\sqrt{-1}\pi^2N^2}{{\rm Im}\tau}\right)^{\frac{11}{20}}
{\rm e}^{\frac{3}{2}\zeta'_{\Delta_1}(0)-3\zeta'_{\Delta_0}(0)}
\frac{1}{2\sqrt{\pi}}\int\limits_{-\infty}^{+\infty}
t^{-\frac{11}{20}}{\rm e}^{-\frac{t^2}{4}}\dd t\\
        \\
\geqq Z_{\varepsilon}(\R^4,\tau)\geqq\\
         \\
\left(\frac{4\sqrt{-1}\:\pi K(\delta)}{{\rm Im}\tau}\right)^\frac{11}{20}
{\rm e}^{\frac{3}{2}\zeta'_{\Delta_1}(0)-3\zeta'_{\Delta_0}(0)}
\frac{1}{2\sqrt{\pi}}\int\limits_{-\infty}^{+\infty}
\left(t+\sqrt{-1}c\sqrt{\frac{4\pi}{{\rm Im}\tau}}\right)^{-\frac{11}{20}}
{\rm e}^{-\frac{t^2}{4}}\:\dd t\:\:.
\label{kulcsegyenlotlenseg2}
\end{array}
\end{equation}

\begin{lemma} There exist constants $0<t_0, \delta_0,R_{\infty}<+\infty$ 
with the following property. For any choice of the complex coupling 
constant (\ref{tau}) satisfying $t_0<{\rm Im}\tau$ 
(with induced vicinity parameter $0<\varepsilon<+\infty$ as in Lemma
\ref{kettosegyenlotlenseglemma} such that (\ref{kulcsegyenlotlenseg2}) holds), 
the left and right hand sides of (\ref{kulcsegyenlotlenseg2}) get equal
with some $\delta_0<\delta(\tau)$ and with every $0<R\leqq R(\tau)$, 
where $R(\tau)<R_{\infty}$. This yields that    
\[Z_{\varepsilon}(\R^4,\tau)=\left(\frac{{\rm Im}\tau}{2\pi^2N^2}
\right)^{-\frac{11}{20}}\:
\frac{2^{-\frac{11}{20}}}{\sqrt{\pi}}\cos\begin{smallmatrix}
\left(\frac{11\pi}{40}\right)\end{smallmatrix}\Gamma
\begin{smallmatrix}\left(\frac{9}{40}\right)\end{smallmatrix}
{\rm e}^{\frac{3}{2}\zeta'_{\Delta_1}(0)-3\zeta'_{\Delta_0}(0)}\]
where $\Gamma$ is Euler's Gamma function.

Moreover the partition function $Z_\varepsilon(\R^4,\tau)$ as calculated here 
depends on $R$, the radius of the conformal compactification $(S^4,g_R)$ 
of the original Euclidean space $(\R^4,\eta)$, only through its determinant 
term ${\rm e}^{\frac{3}{2}\zeta'_{\Delta_1}(0)-3\zeta'_{\Delta_0}(0)}$. 
More precisely if for a given 
$\tau\in\C^+$ two permitted conformal one-point compactifications 
$(S^4,g_{R_i})$ are taken i.e. 
$0<R_1<R_2\leqq R(\tau)<R_{\infty}$ then 
the corresponding partition functions are related by  
$Z^1_\varepsilon(\R^4,\tau)=\left(\frac{R_1}{R_2}
\right)^{\frac{11}{10}}Z^2_\varepsilon(\R^4,\tau)$. 
\label{eredmeny}
\end{lemma}

\begin{proof} It is clear that the 
scissor (\ref{kulcsegyenlotlenseg2}) around the 
partition function closes up if the equation
\begin{equation}
\left(\frac{\pi N^2}{2K(\delta)}\right)^{\frac{11}{20}}=
\frac{\int\limits_{-\infty}^{+\infty}
\left(t+\sqrt{-1}c\sqrt{\frac{4\pi}{{\rm Im}\tau}}\right)^{-\frac{11}{20}}
{\rm e}^{-\frac{t^2}{4}}\:\dd t}{\int\limits_{-\infty}^{+\infty}
t^{-\frac{11}{20}}{\rm e}^{-\frac{t^2}{4}}\dd t}
\label{integral}
\end{equation}
can be solved for some $\delta$ without breaking the inclusion
$C_{\delta}\subset\ca_{\varepsilon, N}(\nabla^0)$ over some $(S^4,g_R)$. 
The right hand side of (\ref{integral}) monotonly grows from $0$ 
to $1$ as $0\leqq{\rm Im}\tau\leqq+\infty$. Likewise via 
$0\leqq K(\delta)\leqq\pi$ the left hand side of (\ref{integral}) 
monotonly decays from $+\infty$ to $\big(\frac{N^2}{2}\big)^{\frac{11}{20}}$ as 
$0\leqq\delta\leqq+\infty$. Assume now that $N<\sqrt{2}$ 
hence $\big(\frac{N^2}{2}\big)^{\frac{11}{20}}<1$. These together imply 
that we can find a constant $0<t_0<+\infty$ such that the right hand side of 
(\ref{integral}), when evaluated at ${\rm Im}\tau=t_0$, is equal to 
$\big(\frac{N^2}{2}\big)^{\frac{11}{20}}$. Likewise we can find another 
constant $0<\delta_0<+\infty$ such that the left hand side of 
(\ref{integral}), when evaluated at the constant $\delta_0$, is equal to $1$. 
It then readily follows that for every $\tau\in\C^+$ satisfying 
$t_0<{\rm Im}\tau$ there exists $\delta_0<\delta (\tau)$ such that 
(\ref{integral}) can be solved. Note that as 
${\rm Im}\tau\rightarrow+\infty$ then $\delta_0\leftarrow\delta(\tau)$
however as $t_0\leftarrow{\rm Im}\tau$ then $\delta(\tau)\rightarrow+\infty$. 
Proceeding further, by shrinking $R$, i.e. conformally 
rescaling $(S^4,g_R)$ with a constant if necessary, we can scale up 
$\lambda_{{\rm min}}$, the smallest eigenvalue of $\Delta_1$, to 
be arbitrary large without affecting the other conformally invariant 
parameters $\varepsilon, N,c$ of 
the theory. Thus for any permitted choice of $\tau\in\C^+$ there exists a 
radius $R(\tau)$ such that working over any $(S^4, g_R)$ obeying 
$0<R\leqq R(\tau)$ we can take $\delta(\tau)$ 
without breaking $0<\delta(\tau)<\sqrt{\lambda_{\rm min}}\:
\min(\eta,N\varepsilon)$ i.e., the inclusion 
$C_{\delta(\tau)}\subset\ca_{\varepsilon, N}(\nabla^0)$ which has been 
used in (\ref{kulcsegyenlotlenseg2}). Again note that as 
${\rm Im}\tau\rightarrow +\infty$ then $R(\tau)\rightarrow
R_{\infty}:=\sup\big\{\mbox{$R\:
\vert\:C_{\delta_0}\subset\ca_{\varepsilon, N}(\nabla^0)$ is valid}\big\}
<+\infty$ but as $t_0\leftarrow{\rm Im}\tau$ then 
$0\leftarrow R(\tau)$. Summarizing, we can consistently solve 
(\ref{integral}) whenever $N<\sqrt{2}$. However this latter 
condition---{\it which is therefore the only but crucial 
condition\footnote{Honestly speaking we
also assume the validity of the {\bf Monotonicity principles} as 
formulated above. However the (in)validity of these assumptions is 
rather related with the more general problem of the existence of a
satisfactory measure theory in infinite dimensions, cf. the Appendix below.}
for our whole method to work here}---is already satisfied for small
$\varepsilon$'s because 
Lemma \ref{kettosegyenlotlenseglemma} makes sure that $N\rightarrow 1$ as 
$\varepsilon\rightarrow 0$. 

Therefore (\ref{kulcsegyenlotlenseg2}) in fact provides us with an equality 
\[Z_{\varepsilon}(\R^4,\tau)=\left(\frac{2\sqrt{-1}\pi^2N^2}{{\rm Im}\tau}
\right)^{\frac{11}{20}}
{\rm e}^{\frac{3}{2}\zeta'_{\Delta_1}(0)-3\zeta'_{\Delta_0}(0)}
\frac{1}{2\sqrt{\pi}}\int\limits_{-\infty}^{+\infty}
t^{-\frac{11}{20}}{\rm e}^{-\frac{t^2}{4}}\dd t\]
and our last task is to evaluate the complex integral here. We can do this 
by executing a counterclockwise rotation of 
the negative part of the integration contour $\R\subset\C$ 
(together with the branch cutting along the negative imaginary axis) about 
the origin towards its positive part; this shows that 
\[\sqrt{-1}\:^\frac{11}{20}\int\limits_{-\infty}^{+\infty}
t^{-\frac{11}{20}}{\rm e}^{-\frac{t^2}{4}}\dd t=\sqrt{-1}\:^\frac{11}{20}
\big(1+(-1)^{-\frac{11}{20}}\big)\int\limits_0^{+\infty}
t^{-\frac{11}{20}}{\rm e}^{-\frac{t^2}{4}}\dd t\] 
with a real integral on the right. Firstly 
$\sqrt{-1}\:^\frac{11}{20}\big(1+\big(\frac{1}{(\sqrt{-1})^2}
\big)^\frac{11}{20}\big)=\sqrt{-1}\:^\frac{11}{20}+\big(\frac{1}{\sqrt{-1}}
\big)^\frac{11}{20}=2\cos\big(\frac{11\pi}{40}\big)$. Secondly the 
substitution $u:=\frac{t^2}{4}$ yields 
$\int_0^{+\infty}t^{-\frac{11}{20}}{\rm e}^{-\frac{t^2}{4}}\dd t 
=2^{-\frac{11}{20}}\Gamma (\frac{9}{40})$ hence the result. 

Concerning the role of the compactification radius, recall that 
$0\leftarrow R(\tau)$ as $t_0\leftarrow{\rm Im}\tau$ consequently 
there exists no overall finite choice for $R$ which could work for every 
permitted value of $\tau$ thus the $R$ dependence of 
$Z_\varepsilon(\R^4,\tau)$, as has been computed here, is unavoidable. 
Nevertheless, since $N$ is conformally 
invariant, $Z_\varepsilon(\R^4,\tau)$ as it stands can depend on $R$ only 
through the functional determinant. If $(S^4,g_{R_i})$ are 
two conformal one-point compactifications of $(\R^4,\eta)$ then obviously
$g_{R_1}=\left(\frac{R_1}{R_2}\right)^2g_{R_2}$ which can be 
regarded as a homothety applied on $g_{R_1}$. Therefore the eigenvalues of 
$\Delta_1$ under this re-sizeing simply change as 
$\lambda_k\mapsto\left(\frac{R_2}{R_1}\right)^2\lambda_k$ i.e. 
coincide with that of the scaled Laplacian 
$\left(\frac{R_2}{R_1}\right)^2\Delta_1$ hence 
$\zeta_{\Delta_1}\mapsto\zeta_{\left(\frac{R_2}{R_1}\right)^2\Delta_1}$. 
Consequently ${\rm e}^{\frac{3}{2}\zeta'_{\Delta_1}(0)-3\zeta'_{\Delta_0}(0)}
\mapsto \left(\frac{R_2}{R_1}\right)^{2\left(-\frac{3}{2}
\left(\zeta_{\Delta_1}(0)-\zeta_{\Delta_0}(0)\right)\right)}
{\rm e}^{\frac{3}{2}\zeta'_{\Delta_1}(0)-3\zeta'_{\Delta_0}(0)}$ 
but we already know that $\frac{3}{2}\left(\zeta_{\Delta_1}(0)-
\zeta_{\Delta_0}(0)\right)=\frac{11}{20}$ hence the asserted scaling of 
$Z_\varepsilon (\R^4,\tau)$ follows.
\end{proof}

\noindent {\it Proof of Theorem \ref{fotetel}}. Putting together the 
contents of Lemmata \ref{kettosegyenlotlenseglemma}, \ref{szamolas} and 
\ref{eredmeny} the result follows.\hspace{0.1in}$\square$


\section{Appendix: There is no good measure in infinite dimensions}
\label{four}


For completeness we recall the following simple but important general 
fact about measures in infinite dimensions. Perhaps this no-go result 
demonstrates in the sharpest way the existence of a deep chasm between 
finite and infinite dimensional integration. We also refer to the excellent 
survey book \cite{joh-lap} to gain a broader picture. 

Let $(X,\mu)$ be any measure space. As a 
very basic demand in measure theory the measure $\mu$ is always assumed to be 
{\it $\sigma$-additive} i.e. $\mu (\sqcup_iA_i)=\sum_i\mu(A_i)$ to hold 
for all countable collection of pairwise disjoint measurable subsets 
$A_1,A_2,.\:.\:.\subset X$. If $X$ admits further structures, further natural 
assumptions can be imposed on a measure. If $X$ can be given the structure of a 
Banach space for instance, then mimicing the properties of the Lebesgue 
measure in finite dimensions, one can further demand $\mu$ to be (i) 
{\it positive} i.e. $0\leqq\mu (U)\leqq +\infty$ for every open 
subset $\emptyset\subseteqq U\subseteqq X$; (ii) {\it locally 
finite} i.e. every point $x\in X$ has an open neighbourhood 
$N_x\subseteqq X$ such that $-\infty<\mu (N_x)<+\infty$; (iii) and finally 
{\it translation invariant} that is for every measurable subset 
$\emptyset\subseteqq A \subseteqq X$ and every vector $x\in X$ the translated 
set $x+A$ is measurable and $\mu (x+A)=\mu (A)$ holds. 

However, as it is well-known, these natural demands conflict each other in 
infinite dimensions:

\begin{theorem}{\rm (cf. e.g. \cite[Theorem 4, p. 359]{gel-vil}, 
or \cite[Theorem 3.1.5]{joh-lap})} 
Let $(X, \Vert\:\cdot\:\Vert)$ be an infinite dimensional, separable Banach 
space. Then the only locally finite and translation invariant Borel measure 
$\mu$ on $X$ is the trivial measure, with $\mu(A) = 0$ for every 
measurable subset $A$. Equivalently, every translation invariant measure that 
is not identically zero assigns infinite measure to all open subsets of $X$.
\end{theorem}

\begin{proof} Take a locally finite, translation 
invariant measure $\mu$ on an infinite dimensional, separable Banach space 
$(X, \Vert\:\cdot\:\Vert)$. Using local finiteness, 
suppose that, for some $\varepsilon > 0$, the open ball $B_\varepsilon(0)
\subset X$ of radius $\varepsilon$ and centered at the origin, has a finite 
$\mu$-measure. Since $X$ is infinite dimensional, there is a countable 
infinite sequence of pairwise disjoint open balls $B_{\frac{\varepsilon}{4}}
(x_i)$ of radius for instance $\frac{\varepsilon}{4}$ and centers $x_i\in X$, 
with all the smaller balls $B_{\frac{\varepsilon}{4}}(x_i)$ with $i=1,2,\dots$ 
contained within the larger ball $B_\varepsilon(0)$. By translation invariance, 
all of the smaller balls have the same measure; since by $\sigma$-additivity 
the absolute value of the sum of these measures is estimated from above by 
$\mu (B_\varepsilon(0))<+\infty$ hence is finite, the smaller balls must all 
have $\mu$-measure zero. Now, since $X$ is separable, it can be covered by a 
countable collection of balls of radius $\frac{\varepsilon}{4}$; since each 
such ball has $\mu$-measure zero, by $\sigma$-additivity again so must the 
whole space $X$. Therefore $\mu$ is the trivial measure.
\end{proof}

\noindent This means that our {\it ad hoc} ``measure'' $\DD a$ used 
for integration in a Hilbert space throughout Sections \ref{three} and 
\ref{four} lacks at least one of the standard properties listed above. We 
already observed in the {\it Remark} after the {\bf Monotonicity principles} 
that our hypothetical $\DD a$ assigns finite measure to certain subsets which 
do not contain open balls at all (like the ``principal axis hypercube'' 
$C_\delta\cap\ker\dd^*$ which is not open in infinite dimensions). This oddity 
might be related with another one too namely that 
it is locally finite for certain open subsets (
like the ball $\ca_{\varepsilon, N}(\nabla^0)\cap\ker\dd^*$ or the 
full Hilbert space $\ker\dd^*$ itself). 
\vspace{0.1in}

\noindent{\bf Acknowledgement}. This paper is dedicated to Karen K. 
Uhlenbeck, the laureate of the 2019 Abel Prize in mathematics. Thanks go to 
P. Vrana for some technical observations. There are no conflicts of 
interest to declare that are relevant to the content of this article. 
The work meets all ethical standards applicable here. 
All the not-referenced results in this work are fully the author's own 
contribution. No funds, grants, or other financial supports were received. 
Data sharing not applicable to this article as no datasets were generated or 
analysed during the corresponding study.

\end{document}